\documentclass[10pt]{article}
\usepackage[colorlinks,
            linkcolor=blue,
            urlcolor=blue,
            anchorcolor=red,
            citecolor=blue
            ]{hyperref}
\usepackage{graphicx}
\usepackage{cite}
\usepackage{subfigure}
\usepackage{epstopdf}
\usepackage{rotating}
\usepackage{amsmath}
\usepackage{amssymb}
\usepackage{float}
\usepackage{color, soul}

\begin{document}
\begin{center}
{\Large \bf Energy dependent chemical potentials of light hadrons and quarks based on transverse momentum spectra and yield ratios of negative to positive particles}

\vskip1.0cm

Xing-Wei He$^{a}$, Feng-Min Wu$^{a,b}$, Hua-Rong Wei$^{c,}${\footnote{E-mail: huarongwei@qq.com;huarongwei@lsu.edu.cn}}, and Bi-Hai Hong$^{c}$

{\small\it $^a$Institute of Theoretical Physics \& State Key Laboratory of Quantum Optics and Quantum Optics Devices, Shanxi University, Taiyuan, Shanxi 030006, People's Republic of China\\
$^{b}$Department of Physics, Zhejiang Sci-Tech University, Hangzhou, Zhejiang 310000, People's Republic of China\\
$^c$Institute of Optoelectronic Technology, Lishui University, Lishui, Zhejiang 323000, People's Republic of China}
\end{center}

\vskip1.0cm

{\bf Abstract:} We describe the transverse momentum spectra or transverse mass spectra of $\pi^\pm$, $K^\pm$, $p$, and $\bar{p}$ produced in central gold-gold (Au-Au), central lead-lead (Pb-Pb), and inelastic proton-proton ($pp$) collisions at different collision energies range from the AGS to LHC by using a two-component (in most cases) Erlang distribution in the framework of multi-source thermal model. The fitting results are consistent with the experimental data, and the final-state yield ratios of negative to positive particles are obtained based on the normalization constants from the above describing the transverse momentum (or mass) spectra. The energy dependent chemical potentials of light hadrons ($\pi$, $K$, and $p$) and quarks ($u$, $d$, and $s$) in central Au-Au, central Pb-Pb, and inelastic $pp$ collisions, are then extracted from the modified yield ratios in which the contributions of strong decay from high-mass resonance and weak decay from heavy flavor hadrons are removed. The study shows that most types of energy dependent chemical potentials decrease with increase of collision energy over a range from the AGS to LHC. The curves of all types of energy dependent chemical potentials, obtained from the fits of yield ratios vs energy, have the maximum at about 3.510 GeV, which possibly is the critical energy of phase transition from a liquid-like hadron state to a gas-like quark state in the collision system. At the top RHIC and LHC, all types of chemical potentials become small and tend to zero at very high energy, which confirms that the high energy collision system possibly changes completely from the liquid-like hadron-dominant state to the gas-like quark-dominant state and the partonic interactions possibly play a dominant role at the LHC.
\\

{\bf Keywords:} transverse momentum spectra, yield ratios of negative to positive particles, chemical potentials of particles, critical end point of phase transition
\\

PACS: 14.65.Bt, 13.85.Hd, 24.10.Pa

\vskip1.0cm

{\section{Introduction}}

The critical energy of phase transition \cite{Xu2014,Grebieszkow2017,Yang2017,Turko(NA61/SHINECollaboration)2018} is important for studying the quantum chromodynamics (QCD) phase diagram \cite{Braun-Munzinger2009,Wambach2007} and the properties of quark-gluon plasma (QGP) \cite{Cleymans2006,Rischke2004,Cabibbo1975}, so more and more scientists devote to finding the critical energy. The experiments performed on the Relativistic Heavy Ion Collider (RHIC) and the Large Hadron Collider (LHC), especially the beam energy scan program at the RHIC, deal with a collision energy range from a few to several tens of GeV \cite{Xu2014,Cleymans2006,Andronic2010,Bellwied2018}, which may contain the energy of the critical end point of hadron-quark phase transition \cite{Xu2014,Grebieszkow2017,Yang2017,Turko(NA61/SHINECollaboration)2018,Lao2019}. The STAR Collaboration found that the critical energy may be or below 19.6 GeV (unless otherwise noted, the energy values presented in this paper are in the center-of-mass coordinate system) \cite{Xu2014}. One study based on yield ratio (the yield ratio of negative to positive particles) and the correlation between collision energy and transverse momentum indicated that the critical energy maybe range from 11.5 GeV to 19.6 GeV  \cite{Xu2014,Abelev(ALICECollaboration)2015,Md.Nasim2015,Abelev(STARCollaboration)2009paper1}, while another study based on yield ratio showed that the critical energy may be about 4 GeV \cite{Lao2019}. Studies about a striking pattern of viscous damping and an excitation function for ($R^{2}_{out}-R^{2}_{side}$) extracted for central collisions indicated the critical energy may be close to 62.4 GeV \cite{Lacey2015,Lacey2014,Adare(PHENIXCollaboration)2014}. It is not hard to see that the value of critical energy has not been determined so far, so finding the critical energy arouses our great interest.

Lattice QCD \cite{Borsanyi2014,Bazavov(HotQCDCollaboration)2014,Soltz2015}, a powerful tool to investigate the QGP matter in high-temperature and high-density system, indicates that the critical end point (CEP) of phase transition on QCD phase diagram is a crossover at small chemical potentials or high collision energies \cite{Stephanov2006,Wygas2018}. So it is important to study baryon chemical potential for finding the CEP on QCD phase diagram. When collisions occur at high energy, especially at RHIC and LHC, the collision system probably creates the QGP matter \cite{Bazavov2012,Adams(STARCollaboration)2005,Grosse-Oetringhaus(theALICECollaboration)2014} where the partonic interactions play an important role, and the baryon chemical potential is small, even close to 1 MeV or zero \cite{Lao2019,Stachel2014,He2019,Gao2018}. While when energy is not very high, the transition from hadron to quark has not yet taken place in the collision system, where the hadronic interactions play an important role \cite{Xu2014,Abelev(ALICECollaboration)2015,Md.Nasim2015,Abelev(STARCollaboration)2009paper1}, and the value of baryon chemical potential is larger. We could predict that the chemical potential corresponding to the CEP should be a inflection point or abrupt change point in chemical potential-energy plane. It is therefore worthwhile to study the trend of chemical potential with energy.

The yield ratio of negative to positive particles is an important quantity in high energy study. Generally, one can get yield ratio by many ways. One way is to directly collect the values of yield ratio from the productive international collaborations, which is a rapid and convenient method. Another way needs the aid of the extracted normalization constant in describing the transverse momentum spectra of negative and positive particles with consistent statistical law, but the workload is huge. In this paper, due to the fact that experiment data of some particles correspond to a narrow range of transverse momentum ($p_{T}$), we adopt the second method to obtain a relatively accurate result for the normalization constant being extracted from a wider range of transverse momentum distribution. In addition, one can extract the yield ratio as a fitting parameter in many models. One can obtain some information about the very hot and dense nuclear matter by yield ratio. For example, one can analyze some statistical thermal models to extract temperature, baryon chemical potential ($\mu_{B}$) at chemical freez-out and so on by describing the ratios of hadron yields, and further to establish the "line of chemical freez-out"\cite{Braun-Munzinger2004paper2} by which one can continue to study QGP, QCD phase transition, and QCD phase diagram\cite{Andronic2006,Braun-Munzinger1998,Cleymans2002,Becattini2006,Andronic2007,Cleymans2006}. Except for baryon chemical potential, one can study the chemical potentials of other hadrons and quarks because these chemical potentials are also important in studying collision system evolution and particle production.

In the present work, we describe the transverse momentum ($p_{T}$) or transverse mass ($m_{T}$) spectra of $\pi^\pm$, $K^\pm$, $p$, and $\bar{p}$ produced in central gold-gold (Au-Au), central lead-lead (Pb-Pb) and inelastic proton-proton ($pp$) collisions in mid-rapidity interval (in most cases) over a center-of-mass energy ($\sqrt{s_{NN}}$) range from the AGS to LHC \cite{Klay(E895Collaboration)2003,Ahle(E-802Collaboration)1998,Ahle(E866andE917Collaborations)2000,Adamczyk(STARCollaboration)2017,Abelev(STARCollaboration)2009paper2,
Adcox(PHENIXCollaboration)2004,Adler(PHENIXCollaboration)2004,Alt(NA49Collaboration)2008,Alt(NA49Collaboration)2006,Afanasiev(NA49Collaboration)2002,
Bearden(NA44Collaboration)2002,Abelev(ALICECollaboration)2013,Aduszkiewicz(NA61/SHINECollaboration)2017,Adare2011,Chatrchyan2012,Sirunyan(CMSCollaboration)2017} by using a two-component (in most cases) Erlang distribution \cite{Liu2008paper1,Liu2008paper2} in the framework of a multi-source thermal model \cite{Liu2008paper2,Liu2004,Liu2008paper3}, and obtain the yield ratios, $\pi^{-}/\pi^{+}$, $K^{-}/K^{+}$, and $\overline{p}/p$, of negative to positive particles according to the extracted normalization constants. The energy dependent chemical potentials of light hadrons ($\pi$, $K$, and $p$) and quarks ($u$, $d$, and $s$) in central Au-Au, central Pb-Pb, and inelastic $pp$ collisions are then extracted from the modified yield ratios in which the contributions of strong decay from high-mass resonance and weak decay from heavy flavor hadrons are removed.

{\section{The model and formulism}}

According to our method, to obtain the normalization constants, we need firstly to describe the $p_{T}$ spectra of $\pi^\pm$, $K^\pm$, $p$, and $\bar{p}$ with a multi-component Erlang distribution \cite{Liu2008paper1,Liu2008paper2} which is in the framework of a multi-source thermal model \cite{Liu2008paper2,Liu2004,Liu2008paper3}. The model assumes that many emission sources are formed in high energy collisions and are classified into a few groups due to the existent of different interacting mechanisms in the collisions and different event samples in experiment measurements. The sources in the same group have the same excitation degree and stay at a common local equilibrium state, which can be described by a Erlang $p_{T}$ distribution. All emission sources in different groups result in the final-state distribution, which can be described by a multi-component Erlang $p_{T}$ distribution.

The multi-component Erlang distribution based on the above multi-source thermal model has the following form. According to thermodynamic system, particles generated from one emission source obey to an exponential distribution of transverse momentum,
\begin{equation}
f_{ij}(p_{tij})=\frac{1}{\langle{p_{tij}}\rangle}\exp{\bigg[-\frac{p_{tij}}{\langle{p_{tij}}\rangle}\bigg]},
\end{equation}
where $p_{tij}$ is the transverse momentum of the $i$-th source in the $j$-th group, and $\langle{p_{tij}}\rangle$ is the mean value of $p_{tij}$. We assume that the source number in the $j$-th group and the transverse momentum of the $m_{j}$ sources are denoted by $m_{j}$ and $p_{T}$, respectively. All the sources in the $j$-th group then result in the folding result of exponential distribution
\begin{equation}
f_{j}(p_{T})=\frac{p_T^{m_{j}-1}}{(m_{j}-1)!\langle{p_{tij}}\rangle^{m_{j}}}\exp{\bigg[-\frac{p_{T}}{\langle{p_{tij}}\rangle}\bigg]},
\end{equation}
which is the normalized Erlang distribution. The contribution of the $l$ group of sources can be expressed as
\begin{equation}
f(p_{T})=\sum_{j=1}^{l}k_{j}f_{j}(p_{T}),
\end{equation}
where $k_{j}$ denotes the relative weight contributed by the $j$th group and meets the normalization $\sum_{j=1}^{l}k_{j}=1$. This is the multi-component Erlang distribution.

In fact, in the present work, we describe the transverse momentum spectra of final-state light flavour particles by using a two-component Erlang distribution, where one component reflects the soft excitation process, while the other one reflects the hard scattering process. The soft process corresponding to low-$p_{T}$  region is regarded as the contribution of the interactions among a few sea quarks and gluons, and the hard process corresponding to high-$p_{T}$ region is regarded as originating from a harder head-on scattering between a few valent quarks. Due to the fact that the experimental data of some particles correspond to a narrow range of $p_{T}$, we adopt one-component Erlang distribution to fit these data.

Some experimental data we collect are about transverse mass distribution, not $p_{T}$ distribution, so we give the transformational relation between $p_{T}$ distribution and $m_{T}$ distribution based on the relation between $p_{T}$ and $m_{T}$ ($m_{T}=\sqrt{p_{T}^{2}+m_{0}^{2}}$, where $m_{0}$ is the rest mass of particle), i.e.
\begin{equation}
\frac{dN}{Ndm_{T}}=\frac{m_{T}}{p_{T}}\frac{dN}{Ndp_{T}}.
\end{equation}

The same as in our previous work \cite{He2019}, in the present work, we only calculate the chemical potentials of some light hadrons ($\pi$, $K$, and $p$), and some light quarks ($u$, $d$, and $s$). For the hadrons containing $c$ or $b$ quark, considering that there is a lack of the experimental data of $p_{T}$ spectra continuously varying with energy, we do not calculate the chemical potentials of the hadrons containing $c$ or $b$ quark, and $c$ and $b$ quarks. In addition, due to the lifetimes of the hadrons containing $t$ quark being too short to measure, we also can not obtain the chemical potentials of the hadrons containing $t$ quark. According to the statistical arguments based on the chemical and thermal equilibrium within the thermal and statistical model \cite{Braun-Munzinger2001}, we can get the relations between antiparticle to particle (negative to positive particle) yield ratios and chemical potentials of hadrons to be \cite{Braun-Munzinger2001,Adler(PHENIXCollaboration)2004,Zhao2015}
\begin{gather}
k_{\pi}=\exp\bigg(-\frac{2\mu_{\pi}}{T_{ch}}\bigg),\notag\\
k_{K}=\exp\bigg(-\frac{2\mu_{K}}{T_{ch}}\bigg),\notag\\
k_{p}=\exp\bigg(-\frac{2\mu_{p}}{T_{ch}}\bigg),
\end{gather}
where $k_{\pi}$, $k_{K}$, and $k_{p}$ denote the yield ratios of antiparticles, $\pi^{-}$, $K^{-}$, and $\overline{p}$, to particles, $\pi^{+}$, $K^{+}$, and $p$, respectively, and $\mu_{\pi}$, $\mu_{K}$, and $\mu_{p}$ represent the chemical potentials of $\pi$, $K$, and $p$, respectively. In addition, $T_{ch}$ represents the chemical freeze-out temperature of interacting system, and can be empirically obtained by the following formula
\begin{equation}
T_{ch}=T_{\lim}\frac{1}{1+\exp[2.60-\ln(\sqrt{s_{NN}})/0.45]}
\end{equation}
within the framework of a statistical thermal model of non-interacting gas particles with the assumption of standard Maxwell-Boltzmann statistics \cite{Cleymans2006,Rischke2004,Andronic2009}, where the `limiting' temperature $T_{\lim}$ is 0.164 GeV, and $\sqrt{s_{NN}}$ is in the unit of GeV \cite{Andronic2009,Kovacs2008}.

Assuming that $\mu_{u}$, $\mu_{d}$, and $\mu_{s}$ represent the chemical potentials of $u$, $d$, and $s$ quarks, respectively, and according to Equation (5) and references \cite{Lao2019,Zhao2015,Arsene2005} under the same value of chemical freeze-out temperature, the yield ratios in terms of quark chemical potentials can be written as
\begin{gather}
k_{\pi}=\exp\bigg[-\frac{(\mu_{u}-\mu_{d})}{T_{ch}}\bigg]\bigg/\exp\bigg[\frac{(\mu_{u}-\mu_{d})}{T_{ch}}\bigg]=\exp\bigg[-\frac{2(\mu_{u}-\mu_{d})}{T_{ch}}\bigg],\notag\\
k_{K}=\exp\bigg[-\frac{(\mu_{u}-\mu_{s})}{T_{ch}}\bigg]\bigg/\exp\bigg[\frac{(\mu_{u}-\mu_{s})}{T_{ch}}\bigg]=\exp\bigg[-\frac{2(\mu_{u}-\mu_{s})}{T_{ch}}\bigg],\notag\\
k_{p}=\exp\bigg[-\frac{(2\mu_{u}+\mu_{d})}{T_{ch}}\bigg]\bigg/\exp\bigg[\frac{(2\mu_{u}+\mu_{d})}{T_{ch}}\bigg]=\exp\bigg[-\frac{2(2\mu_{u}+\mu_{d})}{T_{ch}}\bigg].
\end{gather}

Based on Equations (5) and (7), one can obtain the chemical potentials of hadrons and quarks in terms of yield ratios respectively,
\begin{gather}
\mu_{\pi}=-\frac{1}{2}T_{ch}\cdot\ln(k_{\pi}),\notag\\
\mu_{K}=-\frac{1}{2}T_{ch}\cdot\ln(k_{K}),\notag\\
\mu_{p}=-\frac{1}{2}T_{ch}\cdot\ln(k_{p}),
\end{gather}
and
\begin{gather}
\mu_{u}=-\frac{1}{6}T_{ch}\cdot\ln(k_{\pi}\cdot k_{p}),\notag\\
\mu_{d}=-\frac{1}{6}T_{ch}\cdot\ln(k_{\pi}^{-2}\cdot k_{p}),\notag\\
\mu_{s}=-\frac{1}{6}T_{ch}\cdot\ln(k_{\pi}\cdot k_{K}^{-3}\cdot
k_{p}).
\end{gather}

In the present work, by describing the $p_{T}$ (or $m_{T}$ ) spectra of some light particles, $\pi^\pm$, $K^\pm$, $p$, and $\bar{p}$ in central Au-Au, central Pb-Pb, and inelastic $pp$ collisions in mid-rapidity interval at collision energy from the AGS to LHC with a two-component (in most cases) Erlang distribution, we obtain the yield ratios, $\pi^{-}/\pi^{+}$, $K^{-}/K^{+}$, and $\overline{p}/p$ based on the extracted normalization constants, and the chemical potentials of light hadrons ($\pi$, $K$, and $p$) and light quarks ($u$, $d$, and $s$) based on the yield ratios modified by removing the contributions of strong decay from high-mass resonance and weak decay from heavy flavor hadrons. Then the dependencies of chemical potentials on $\sqrt{s_{NN}}$ are analyzed.

{\section{Results and discussion}}

Figure 1 shows the transverse mass distributions of $\pi^{\pm}$ [Figures 1(a) and 1(b)] and $K^{\pm}$ [Figures 1(c) and 1(d)] produced in central (0--5\%) Au-Au collisions at mid-rapidity in the center-of-mass energy ($\sqrt{s_{NN}}$) range from  2.67 to 4.84 GeV, where $dN/dy$ on axis denote the rapidity density. The experimental data represented by different kinds of symbols were measured by the E895 Collaboration \cite{Klay(E895Collaboration)2003} for $\pi^{\pm}$ at 2.67, 3.31, 3.81, and 4.28 GeV, and the E866 and E917 Collaborations \cite{Ahle(E-802Collaboration)1998,Ahle(E866andE917Collaborations)2000} for $K^{\pm}$ at 3.31, 3.81, 4.28 and 4.84 GeV. The data at each energy are scaled by suitable factors for clarity. The plotted errors bars include both statistical and systematic uncertainties for $\pi^{\pm}$ and only statistical uncertainty for $K^{\pm}$. The solid curves are our results calculated by using the two-component Erlang distribution. The values of free parameters ($m_{1}$, $p_{ti1}$, $k_{1}$, $m_{2}$, and $p_{ti2}$), normalization constant ($N_0$), and $\chi^2$ per degree of freedom ($\chi^2$/dof), and p-values corresponding to the two-component Erlang distribution are listed in Table 1, where the normalization constant is for comparison between curve and data. Here, $\chi^2$ is calculated according to the following formula of
\begin{gather}
\chi^2= \sum{\frac{(N_{i}^{exp}-N_{i}^{theo})^{2}}{\sigma^{2}}}
\end{gather}
where $N_{i}^{exp}$, $N_{i}^{theo}$ and $\sigma$ denote experimental value, theoretical value, and error value, respectively. One can see that the two-component Erlang distribution can well describe the experimental data of the considered particles in Au-Au collisions at the AGS. The values of $m_{2}$ corresponding to high-$p_{T}$ region for different particles are 2, which reflects that the hard process origins from a hard head-on scattering between two valent quarks, while the values of $m_{1}$ corresponding to low-$p_{T}$ region for different particles are 3, which reflects that the soft process origins from the interaction among a few sea quarks and gluons. The values of weight factor $k_1$ of soft excitation process are more than 50\%, which shows that soft excitation is the main excitation process, and the normalization constants $N_0$ increases with increase of energy. It should be noted that the particle yield ratio is represented by $N_0$ from the spectrum of negative or positive particles. The relative value of $N_0$ is enough to obtain the particle yield ratio.

Figure 2 presents the transverse momentum spectra of $\pi^\pm$, $K^\pm$, $p$, and $\bar{p}$ in central (0--5\%) Au-Au collisions at center-of-mass energy 7.7 [Figures 2(a) and 2(d)], 11.5 (Figures 2(b) and 2(e)), and 19.6 [Figures 2(c) and 2(f)] GeV. The symbols represent the experimental data recorded by the STAR Collaboration in the mid-rapidity range $|y|<0.1$ \cite{Adamczyk(STARCollaboration)2017}. The uncertainties are statistical and systematic added in quadrature. The curves are our results fitted by using the two-component Erlang distribution. The values of $m_{1}$, $p_{ti1}$, $k_{1}$, $m_{2}$, $p_{ti2}$, $N_0$, and $\chi^2$/dof, and p-values corresponding to the two-component Erlang distribution are given in Table 1. It is not hard to see that the experimental data can be well fitted by the two-component Erlang distribution. Similarly, the values of $m_{2}$ are 2, and the values of $m_{1}$ are 2, 3, and 4. The values of weight factor $k_1$ are more than 50\%, and $N_0$ in most cases increases with increase of collision energy.

Figure 3 gives the same as Figure 2 but for Au-Au collisions at 27 [Figures 3(a) and 3(d)], 39 [Figures 3(b) and 3(e)], and 62.4 [Figures 3(c) and 3(f)] GeV. All the experimental data were recorded by the STAR Collaboration \cite{Adamczyk(STARCollaboration)2017,Abelev(STARCollaboration)2009paper2}. The results calculated by using the two-component Erlang distribution are shown in the solid curves, where the values of corresponding free parameters, normalization constant, and $\chi^2$/dof, and p-values are shown in Table 2. Obviously, the calculation results by the two-component Erlang distribution are in good agreement with the experimental data of the considered particles in Au-Au collisions. Once more, the values of $m_{2}$ are 2, and the values of $m_{1}$ are 2, 3, and 4. The values of weight factor $k_1$ are more than 50\%, and $N_0$ in most cases increases with increase of collision energy.

The $p_{T}$ spectra of $\pi^\pm$, $K^\pm$, $p$, and $\bar{p}$ in central (0--5\%) Au-Au collisions at 130 [Figures 4(a) and 4(c)], 200 [Figures 4(b) and 4(d)] GeV are displayed in Figure 4. The symbols also denote the experimental data recorded by the PHENIX Collaboration \cite{Adcox(PHENIXCollaboration)2004,Adler(PHENIXCollaboration)2004}. The data for each type of particle are divided by suitable factors for clarity. The error bars indicate the combined uncorrelated statistical and systematic uncertainties for 130 GeV, and are statistical only for 200 GeV. The curves are the two-component Erlang model fits to the spectra. The values of all free parameters, normalization constant, and $\chi^2$/dof, and p-values corresponding to the two-component Erlang distribution are listed in Table 2. Similarly, our calculation results with the two-component Erlang model are consistent with the experimental data. The values of $m_{2}$ are 2, and the values of $m_{1}$ are 2, 3, and 4. The values of weight factor $k_1$ are more than 50\%, and $N_0$ in most cases increases with increase of collision energy.

\begin{figure}[H]
\hskip-0.0cm {\centering
\includegraphics[width=14.0cm]{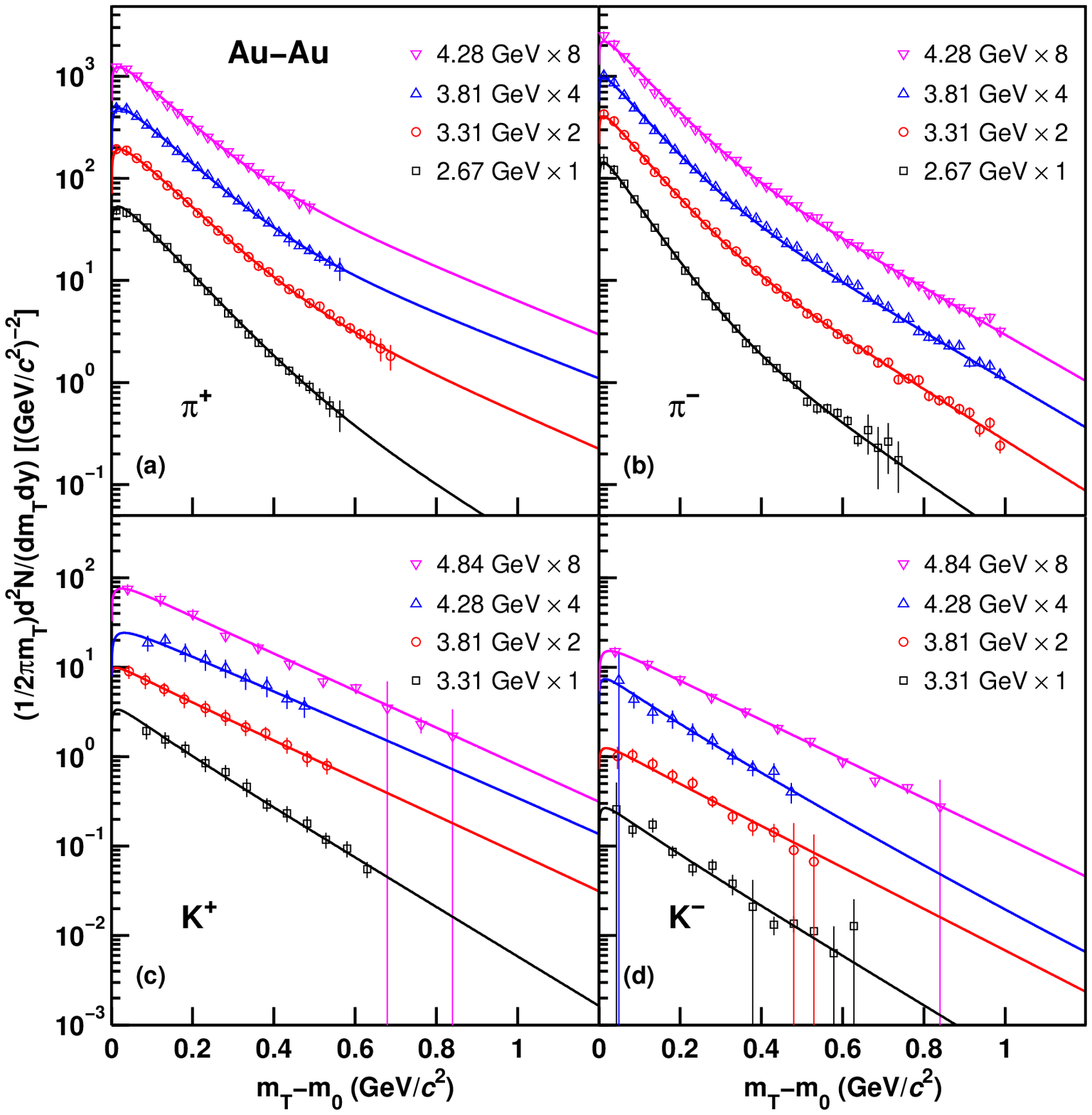}}
\vskip-0.18cm Figure 1. Transverse mass spectra for positive ($\pi^+$, $K^+$) and negative ($\pi^-$, $K^-$) particles produced in central Au-Au collisions at mid-rapidity over a energy range from 2.67 to 4.84 GeV. The experimental data represented by the symbols are measured by the E895 Collaboration \cite{Klay(E895Collaboration)2003} for $\pi^{\pm}$ at 2.67, 3.31, 3.81, and 4.28 GeV, and the E866 and E917 Collaborations \cite{Ahle(E-802Collaboration)1998,Ahle(E866andE917Collaborations)2000} for $K^{\pm}$ at 3.31, 3.81, 4.28 and 4.84 GeV. The data at each energy are scaled by successive powers of 2 for clarity. The plotted errors bars include both statistical and systematic uncertainties for $\pi^{\pm}$ and only statistical uncertainty for $K^{\pm}$. The solid curves are our results calculated by using the two-component Erlang distribution.
\end{figure}

\begin{figure}[H]
\hskip-0.0cm {\centering
\includegraphics[width=16.0cm]{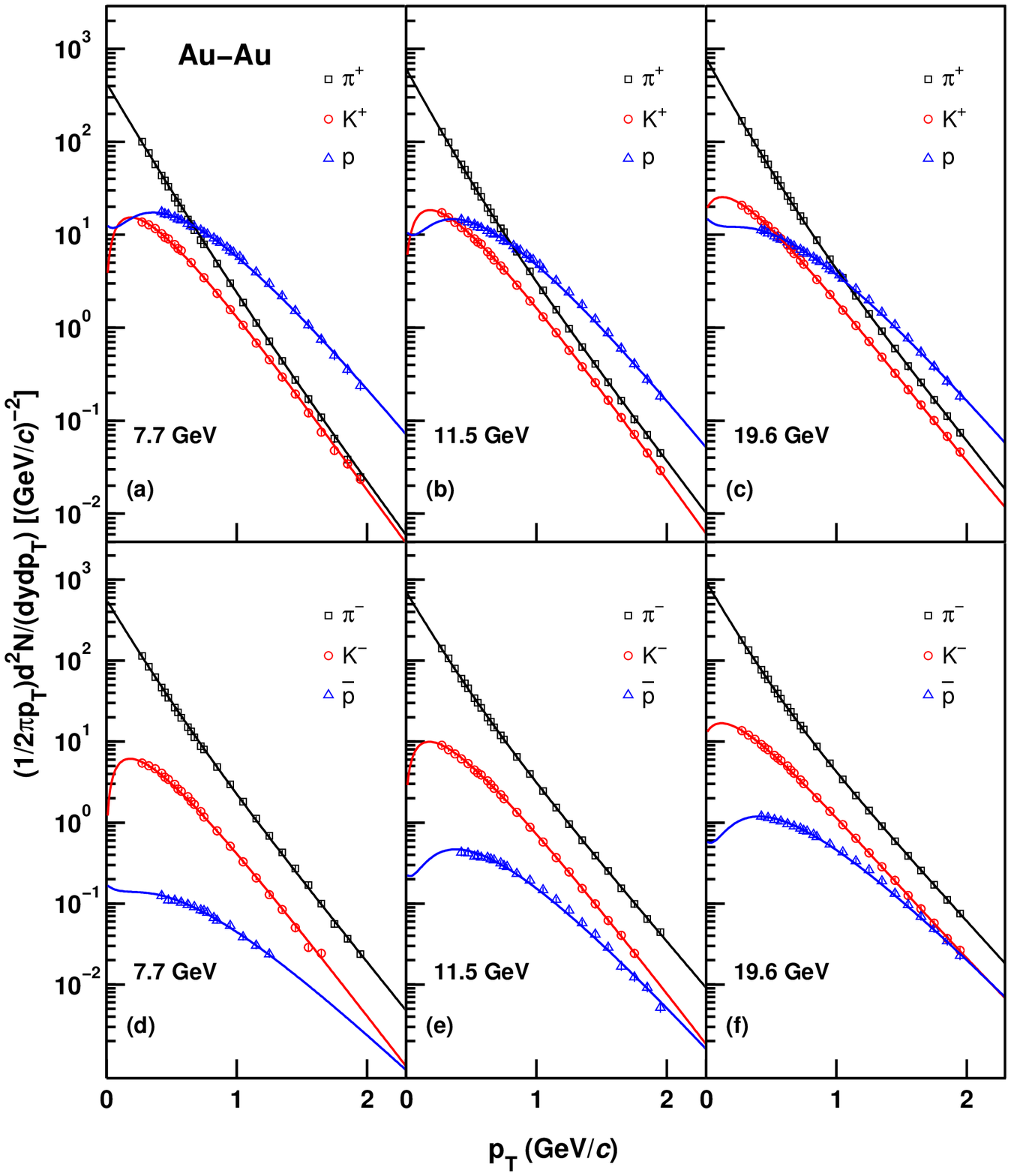}}
\vskip-0.18cm  Figure 2. Mid-rapidity transverse momentum spectra for positive ($\pi^+$, $K^+$, $p$) and negative ($\pi^-$, $K^-$, $\bar{p}$) particles produced in central Au-Au collisions at 7.7 [Figures 2(a) and 2(d)], 11.5 [Figures 2(b) and 2(e)], and 19.6 [Figures 2(c) and 2(f)] GeV. The symbols represent the experimental data recorded by the STAR Collaboration \cite{Adamczyk(STARCollaboration)2017}. The errors are the combined statistical and systematic ones, and the curves are our results by the two-component Erlang distribution.
\end{figure}
\newpage
{\scriptsize {Table 1. Values of free parameters, normalization constant, and $\chi^2$/dof, and p-values corresponding to two-component Erlang $p_T$ (or $m_T$) distribution for Au-Au collisions in Figures 1 and 2.
{%
\begin{center}
\begin{tabular}{ccccccccccc}
\hline
\hline
Figure  & $\sqrt{s_{NN}}$ & Particle & $m_1$ &$<p_{ti1}>$ & $k_1$ & $m_2$ &$<p_{ti2}>$ & $N_0$& $\chi^2$/dof&p-value \\
        &  (GeV)          &          &       & (GeV/c)    &       &       & (GeV/c)    &      &        &      \\
\hline
Figure 1 (a) & 2.67 & $\pi^{+}$      & 3 & 0.078 $\pm$ 0.001 & 0.74 $\pm$ 0.01 & 2 & 0.182 $\pm$ 0.012 & 12.051 $\pm$ 0.365 & 7.043/17& 0.983\\
Figure 1 (b) &      & $\pi^{-}$      & 3 & 0.061 $\pm$ 0.001 & 0.74 $\pm$ 0.02 & 2 & 0.163 $\pm$ 0.010 & 20.811 $\pm$ 0.365 & 11.690/24& 0.983\\
\hline
Figure 1 (a) & 3.31 & $\pi^{+}$      & 3 & 0.083 $\pm$ 0.002 & 0.68 $\pm$ 0.02 & 2 & 0.251 $\pm$ 0.010 & 27.819 $\pm$ 0.747 & 3.768/22& 1\\
Figure 1 (c) &      & $K^{+}$        & 3 & 0.149 $\pm$ 0.003 & 0.91 $\pm$ 0.06 & 2 & 0.162 $\pm$ 0.020 &  2.344 $\pm$ 0.172 & 2.144/6& 0.906\\
Figure 1 (b) &      & $\pi^{-}$      & 3 & 0.067 $\pm$ 0.001 & 0.58 $\pm$ 0.01 & 2 & 0.176 $\pm$ 0.004 & 38.257 $\pm$ 0.777 & 19.052/34& 0.982\\
Figure 1 (d) &      & $K^{-}$        & 3 & 0.149 $\pm$ 0.008 & 0.91 $\pm$ 0.06 & 2 & 0.162 $\pm$ 0.060 &  0.186 $\pm$ 0.017 & 8.857/7& 0.263\\
\hline
Figure 1 (a) & 3.81 & $\pi^{+}$      & 3 & 0.084 $\pm$ 0.002 & 0.60 $\pm$ 0.02 & 2 & 0.280 $\pm$ 0.010 & 39.084 $\pm$ 0.770 & 2.654/17& 1\\
Figure 1 (c) &      & $K^{+}$        & 3 & 0.190 $\pm$ 0.011 & 0.80 $\pm$ 0.06 & 2 & 0.202 $\pm$ 0.023 &  4.829 $\pm$ 0.285 & 0.431/5& 0.994\\
Figure 1 (b) &      & $\pi^{-}$      & 3 & 0.068 $\pm$ 0.001 & 0.51 $\pm$ 0.01 & 2 & 0.187 $\pm$ 0.004 & 50.197 $\pm$ 0.797 & 36.777/34& 0.341\\
Figure 1 (d) &      & $K^{-}$        & 3 & 0.174 $\pm$ 0.007 & 0.88 $\pm$ 0.06 & 2 & 0.185 $\pm$ 0.026 &  0.572 $\pm$ 0.007 & 3.598/5& 0.608\\
\hline
Figure 1 (a) & 4.28 & $\pi^{+}$      & 3 & 0.081 $\pm$ 0.003 & 0.51 $\pm$ 0.02 & 2 & 0.267 $\pm$ 0.016 & 49.619 $\pm$ 1.103 & 5.465/14& 0.978\\
Figure 1 (c) &      & $K^{+}$        & 3 & 0.197 $\pm$ 0.011 & 0.94 $\pm$ 0.04 & 2 & 0.235 $\pm$ 0.030 &  7.717 $\pm$ 0.031 & 1.304/3& 0.728\\
Figure 1 (b) &      & $\pi^{-}$      & 3 & 0.072 $\pm$ 0.003 & 0.51 $\pm$ 0.02 & 2 & 0.192 $\pm$ 0.002 & 62.627 $\pm$ 0.744 & 40.248/34& 0.213\\
Figure 1 (d) &      & $K^{-}$        & 3 & 0.150 $\pm$ 0.009 & 0.80 $\pm$ 0.07 & 2 & 0.235 $\pm$ 0.032 &  1.355 $\pm$ 0.107 & 1.738/4& 0.784\\
\hline
Figure 1 (c) & 4.84 & $K^{+}$        & 3 & 0.187 $\pm$ 0.006 & 0.87 $\pm$ 0.07 & 2 & 0.267 $\pm$ 0.040 & 10.798 $\pm$ 0.383 & 5.328/5& 0.377\\
Figure 1 (d) &      & $K^{-}$        & 3 & 0.180 $\pm$ 0.003 & 0.91 $\pm$ 0.06 & 2 & 0.247 $\pm$ 0.030 &  2.025 $\pm$ 0.069 & 5.463/5& 0.362\\
\hline
Figure 2 (a) & 7.7  & $\pi^{+}$      & 2 & 0.172 $\pm$ 0.004 & 0.63 $\pm$ 0.06 & 2 & 0.233 $\pm$ 0.004 & 96.122 $\pm$ 3.326 & 13.970/20& 0.832\\
             &      & $K^{+}$        & 3 & 0.197 $\pm$ 0.003 & 0.93 $\pm$ 0.07 & 2 & 0.300 $\pm$ 0.033 & 20.070 $\pm$ 0.662 & 6.218/17& 0.991\\
             &      & $p$            & 4 & 0.215 $\pm$ 0.003 & 0.89 $\pm$ 0.08 & 2 & 0.270 $\pm$ 0.054 & 52.211 $\pm$ 2.203 & 5.070/23& 1\\
Figure 2 (d) &      & $\pi^{-}$      & 2 & 0.149 $\pm$ 0.006 & 0.52 $\pm$ 0.03 & 2 & 0.219 $\pm$ 0.003 &107.122 $\pm$ 3.578 & 9.297/20& 0.979\\
             &      & $K^{-}$        & 3 & 0.186 $\pm$ 0.003 & 0.92 $\pm$ 0.08 & 2 & 0.285 $\pm$ 0.057 &  7.208 $\pm$ 0.268 & 14.213/17& 0.652\\
             &      & $\overline{p}$ & 4 & 0.232 $\pm$ 0.013 & 0.71 $\pm$ 0.14 & 2 & 0.334 $\pm$ 0.066 &  0.412 $\pm$ 0.020 & 2.045/9& 0.991\\
\hline
Figure 2 (b) & 11.5 & $\pi^{+}$      & 2 & 0.153 $\pm$ 0.007 & 0.52 $\pm$ 0.04 & 2 & 0.236 $\pm$ 0.003 &125.208 $\pm$ 4.583 & 2.186/20& 1\\
             &      & $K^{+}$        & 3 & 0.201 $\pm$ 0.003 & 0.87 $\pm$ 0.13 & 2 & 0.262 $\pm$ 0.048 & 24.436 $\pm$ 0.718 & 1.486/19& 1\\
             &      & $p$            & 4 & 0.211 $\pm$ 0.003 & 0.90 $\pm$ 0.08 & 2 & 0.244 $\pm$ 0.048 & 42.924 $\pm$ 1.871 & 7.386/22& 0.998\\
Figure 2 (e) &      & $\pi^{-}$      & 2 & 0.146 $\pm$ 0.008 & 0.51 $\pm$ 0.04 & 2 & 0.230 $\pm$ 0.003 &135.170 $\pm$ 5.785 & 1.593/20& 1\\
             &      & $K^{-}$        & 3 & 0.191 $\pm$ 0.003 & 0.92 $\pm$ 0.08 & 2 & 0.221 $\pm$ 0.044 & 12.017 $\pm$ 0.382 & 1.338/17& 1\\
             &      & $\overline{p}$ & 4 & 0.209 $\pm$ 0.004 & 0.92 $\pm$ 0.08 & 2 & 0.234 $\pm$ 0.046 &  1.374 $\pm$ 0.059 & 17.977/17& 0.390\\
\hline
Figure 2 (c) & 19.6 & $\pi^{+}$      & 2 & 0.156 $\pm$ 0.008 & 0.57 $\pm$ 0.05 & 2 & 0.249 $\pm$ 0.005 &165.077 $\pm$ 7.957 & 0.737/20& 1\\
             &      & $K^{+}$        & 3 & 0.198 $\pm$ 0.005 & 0.67 $\pm$ 0.13 & 2 & 0.294 $\pm$ 0.009 & 29.706 $\pm$ 0.879 & 1.316/20& 1\\
             &      & $p$            & 4 & 0.224 $\pm$ 0.004 & 0.79 $\pm$ 0.07 & 2 & 0.278 $\pm$ 0.040 & 34.690 $\pm$ 1.374 & 4.885/23& 1\\
Figure 2 (f) &      & $\pi^{-}$      & 2 & 0.145 $\pm$ 0.008 & 0.56 $\pm$ 0.04 & 2 & 0.246 $\pm$ 0.004 &176.077 $\pm$ 8.381 & 0.682/20& 1\\
             &      & $K^{-}$        & 3 & 0.192 $\pm$ 0.004 & 0.64 $\pm$ 0.07 & 2 & 0.292 $\pm$ 0.007 & 18.620 $\pm$ 0.585 & 2.950/20& 1\\
             &      & $\overline{p}$ & 4 & 0.222 $\pm$ 0.004 & 0.93 $\pm$ 0.07 & 2 & 0.247 $\pm$ 0.049 &  3.937 $\pm$ 0.169 & 7.144/16& 0.970\\
\hline
\end{tabular}
\end{center}
}} }
\begin{figure}[H]
\hskip-0.0cm {\centering
\includegraphics[width=16.0cm]{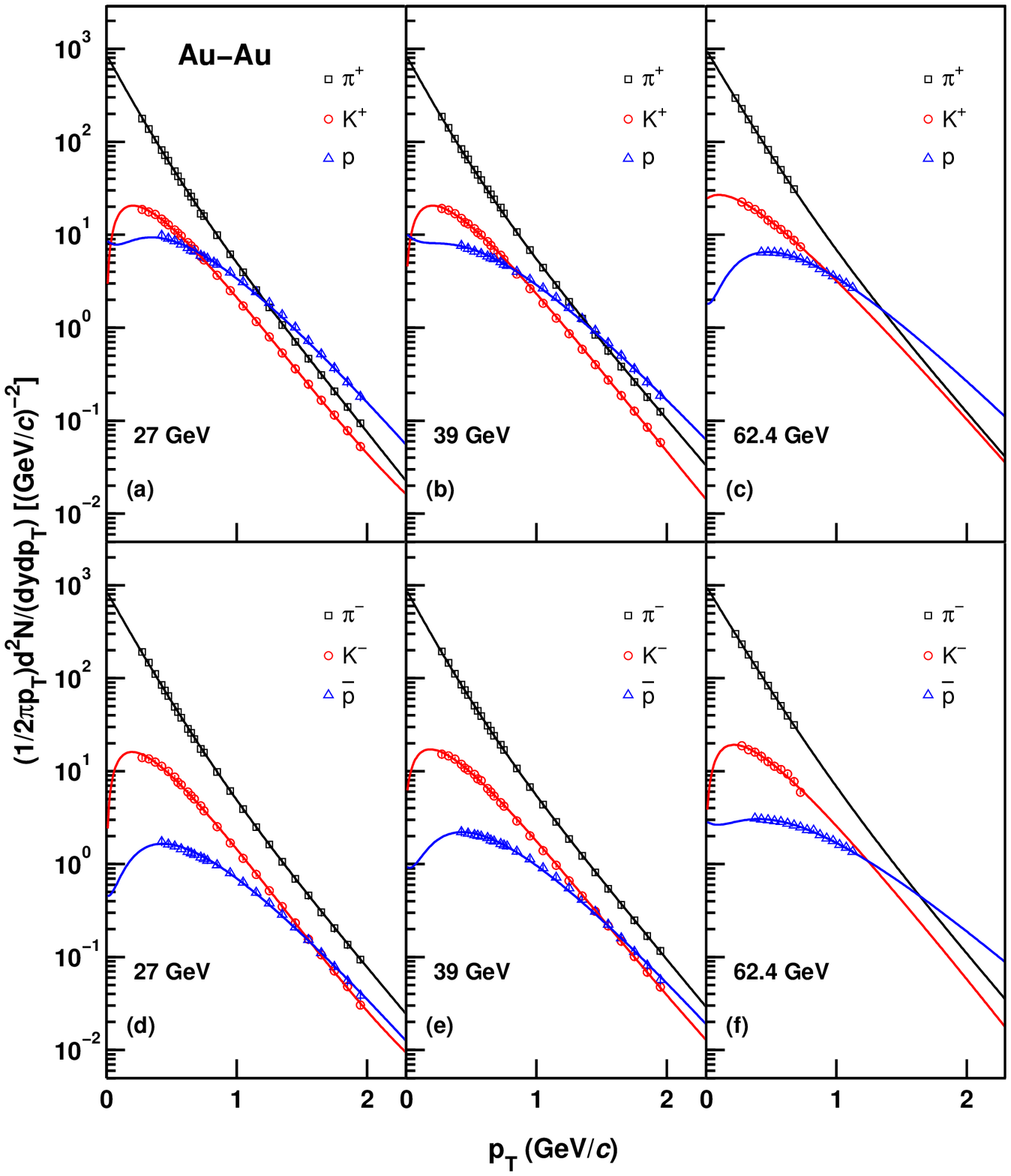}}
\vskip-0.18cm  Figure 3. Same as Figure 2 but for Au-Au collisions at 27 [Figures 3(a) and 3(d)], 39 [Figures 3(b) and 3(e)], and 62.4 [Figures 3(c) and 3(f)] GeV.
\end{figure}

\begin{figure}[H]
\hskip-0.0cm {\centering
\includegraphics[width=16.0cm]{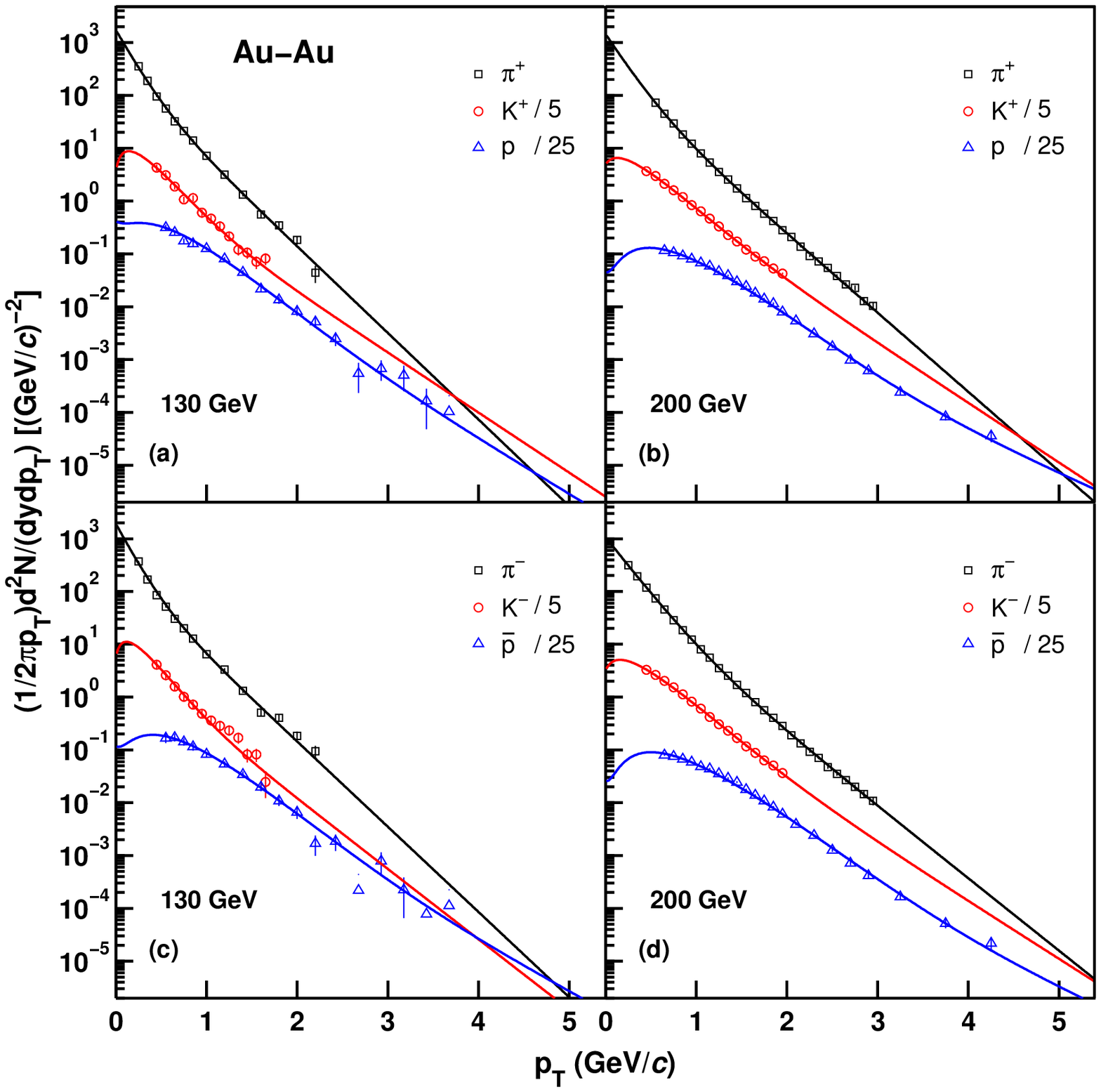}}
\vskip-0.18cm  Figure 4. Same as Figure 2 but for Au-Au collisions at 130 [Figures 4(a) and 4(c)], 200 (Figures 4(b) and 4(d)] GeV.
\end{figure}

\newpage
{\scriptsize {Table 2. Values of free parameters, normalization constant, and $\chi^2$/dof, and p-values corresponding to two-component Erlang $p_T$ (or $m_T$) distribution for Au-Au collisions in Figures 3 and 4.
{%
\begin{center}
\begin{tabular}{ccccccccccc}
\hline
\hline
Figure  & $\sqrt{s_{NN}}$ & Particle & $m_1$ &$<p_{ti1}>$ & $k_1$ & $m_2$ &$<p_{ti2}>$ & $N_0$& $\chi^2$/dof& p-value \\
        &  (GeV)          &          &       & (GeV/c)    &       &       & (GeV/c)    &      &             &         \\
\hline
Figure 3 (a) & 27   & $\pi^{+}$      & 2 & 0.152 $\pm$ 0.007 & 0.51 $\pm$ 0.05 & 2 & 0.249 $\pm$ 0.003 &182.402 $\pm$ 6.202 & 2.449/20 & 1\\
             &      & $K^{+}$        & 3 & 0.205 $\pm$ 0.004 & 0.97 $\pm$ 0.02 & 2 & 0.575 $\pm$ 0.115 & 29.993 $\pm$ 1.046 & 2.393/20 & 1\\
             &      & $p$            & 4 & 0.226 $\pm$ 0.005 & 0.85 $\pm$ 0.12 & 2 & 0.284 $\pm$ 0.056 & 30.191 $\pm$ 1.165 & 4.566/17 & 0.999\\
Figure 3 (d) &      & $\pi^{-}$      & 2 & 0.164 $\pm$ 0.006 & 0.66 $\pm$ 0.04 & 2 & 0.264 $\pm$ 0.004 &186.402 $\pm$ 6.710 & 2.325/20 & 1\\
             &      & $K^{-}$        & 3 & 0.198 $\pm$ 0.002 & 0.97 $\pm$ 0.01 & 2 & 0.531 $\pm$ 0.052 & 21.872 $\pm$ 0.819 & 8.210/19 & 0.984\\
             &      & $\overline{p}$ & 4 & 0.228 $\pm$ 0.004 & 0.92 $\pm$ 0.08 & 2 & 0.343 $\pm$ 0.068 &  5.877 $\pm$ 0.239 & 5.235/16 & 0.994\\
\hline
Figure 3 (b) & 39   & $\pi^{+}$      & 2 & 0.155 $\pm$ 0.008 & 0.54 $\pm$ 0.05 & 2 & 0.265 $\pm$ 0.004 &185.159 $\pm$ 7.258 & 3.265/20 & 1\\
             &      & $K^{+}$        & 3 & 0.211 $\pm$ 0.003 & 0.94 $\pm$ 0.06 & 2 & 0.359 $\pm$ 0.045 & 31.219 $\pm$ 1.024 & 5.650/20 & 0.999\\
             &      & $p$            & 4 & 0.239 $\pm$ 0.005 & 0.79 $\pm$ 0.09 & 2 & 0.293 $\pm$ 0.055 & 26.115 $\pm$ 1.081 & 3.078/16 & 1\\
Figure 3 (e) &      & $\pi^{-}$      & 2 & 0.153 $\pm$ 0.010 & 0.53 $\pm$ 0.04 & 2 & 0.258 $\pm$ 0.004 &191.409 $\pm$ 7.159 & 1.088/20 & 1\\
             &      & $K^{-}$        & 3 & 0.206 $\pm$ 0.003 & 0.86 $\pm$ 0.06 & 2 & 0.352 $\pm$ 0.020 & 24.658 $\pm$ 0.715 & 5.446/20 & 0.999\\
             &      & $\overline{p}$ & 4 & 0.233 $\pm$ 0.004 & 0.92 $\pm$ 0.08 & 2 & 0.280 $\pm$ 0.056 &  8.086 $\pm$ 0.319 & 5.566/17 & 0.996\\
\hline
Figure 3 (c) & 62.4 & $\pi^{+}$      & 2 & 0.172 $\pm$ 0.003 & 0.65 $\pm$ 0.05 & 2 & 0.274 $\pm$ 0.012 &232.461 $\pm$ 1.720 & 0.261/4 & 0.992\\
             &      & $K^{+}$        & 3 & 0.246 $\pm$ 0.004 & 0.75 $\pm$ 0.04 & 2 & 0.256 $\pm$ 0.015 & 39.598 $\pm$ 2.032 & 0.463/4 & 0.977\\
             &      & $p$            & 4 & 0.252 $\pm$ 0.001 & 0.92 $\pm$ 0.03 & 2 & 0.436 $\pm$ 0.027 & 28.457 $\pm$ 0.154 & 2.891/9 & 0.968\\
Figure 3 (f) &      & $\pi^{-}$      & 2 & 0.175 $\pm$ 0.002 & 0.67 $\pm$ 0.05 & 2 & 0.269 $\pm$ 0.008 &234.954 $\pm$ 1.269 & 0.825/4 & 0.935\\
             &      & $K^{-}$        & 3 & 0.223 $\pm$ 0.015 & 0.90 $\pm$ 0.10 & 2 & 0.239 $\pm$ 0.047 & 32.071 $\pm$ 1.844 & 6.873/4 & 0.143\\
             &      & $\overline{p}$ & 4 & 0.281 $\pm$ 0.002 & 0.77 $\pm$ 0.03 & 2 & 0.435 $\pm$ 0.010 & 15.011 $\pm$ 0.117 &11.697/10 & 0.306\\
\hline
Figure 4 (a) & 130  & $\pi^{+}$      & 2 & 0.128 $\pm$ 0.003 & 0.58 $\pm$ 0.02 & 2 & 0.262 $\pm$ 0.020 &288.154 $\pm$ 9.350 & 19.320/8 & 0.013\\
             &      & $K^{+}$        & 3 & 0.166 $\pm$ 0.007 & 0.65 $\pm$ 0.03 & 2 & 0.382 $\pm$ 0.030 & 46.172 $\pm$ 1.883 & 11.031/7 & 0.137\\
             &      & $p$            & 4 & 0.224 $\pm$ 0.008 & 0.61 $\pm$ 0.04 & 2 & 0.420 $\pm$ 0.015 & 29.451 $\pm$ 0.552 & 12.703/11 & 0.313\\
Figure 4 (c) &      & $\pi^{-}$      & 2 & 0.115 $\pm$ 0.010 & 0.60 $\pm$ 0.02 & 2 & 0.269 $\pm$ 0.012 &291.470 $\pm$ 9.100 & 25.039/8 & 0.002\\
             &      & $K^{-}$        & 3 & 0.143 $\pm$ 0.007 & 0.59 $\pm$ 0.04 & 2 & 0.326 $\pm$ 0.016 & 44.407 $\pm$ 1.933 & 11.880/7 & 0.104\\
             &      & $\overline{p}$ & 4 & 0.240 $\pm$ 0.007 & 0.79 $\pm$ 0.04 & 2 & 0.466 $\pm$ 0.030 & 18.813 $\pm$ 0.644 & 17.215/11 & 0.102\\
\hline
Figure 4 (b) & 200  & $\pi^{+}$      & 2 & 0.162 $\pm$ 0.008 & 0.62 $\pm$ 0.02 & 2 & 0.291 $\pm$ 0.004 &314.469 $\pm$ 3.500 & 37.817/19 & 0.006\\
             &      & $K^{+}$        & 3 & 0.208 $\pm$ 0.003 & 0.60 $\pm$ 0.03 & 2 & 0.409 $\pm$ 0.006 & 47.468 $\pm$ 0.793 & 28.553/10 & 0.002\\
             &      & $p$            & 4 & 0.266 $\pm$ 0.003 & 0.92 $\pm$ 0.02 & 2 & 0.568 $\pm$ 0.007 & 15.345 $\pm$ 0.266 & 19.924/16 & 0.224\\
Figure 4 (d) &      & $\pi^{-}$      & 2 & 0.179 $\pm$ 0.004 & 0.65 $\pm$ 0.01 & 2 & 0.297 $\pm$ 0.001 &293.451 $\pm$ 2.500 & 38.653/22 & 0.016\\
             &      & $K^{-}$        & 3 & 0.223 $\pm$ 0.002 & 0.79 $\pm$ 0.01 & 2 & 0.480 $\pm$ 0.007 & 42.471 $\pm$ 0.456 & 9.724/10 & 0.465\\
             &      & $\overline{p}$ & 4 & 0.270 $\pm$ 0.003 & 0.98 $\pm$ 0.02 & 2 & 0.650 $\pm$ 0.006 & 11.195 $\pm$ 0.225 & 26.068/16 & 0.053\\
\hline
\end{tabular}
\end{center}
}} }

Figure 5 exhibits the $m_{T}$ spectra of $\pi^\pm$ at $0<y<0.2$, $K^\pm$ at $|y|<0.1$, $p$, and $\bar{p}$ produced in central Pb-Pb collisions at 6.3 [Figures 5(a) and 5(d)], 7.7 [Figures 5(b) and 5(e)], and 8.8 [Figures 5(c) and 5(f)] GeV. The experimental data, represented by symbols, were taken by the NA49 Collaboration \cite{Alt(NA49Collaboration)2008,Alt(NA49Collaboration)2006,Afanasiev(NA49Collaboration)2002}, where $p$ and $\bar{p}$ were done near mid-rapidity and covered the rapidity intervals of $1.5 < y < 2.2$ ($y_{c.m.} = 1.88$) for 6.3 GeV , $1.6 < y < 2.3$ ($y_{c.m.} = 2.08$) for 7.7 GeV, and $1.9 < y < 2.3$ ($y_{c.m.} = 2.22$) for 8.8 GeV. The error bars on the spectra points are statistical only. The curves are fits of two-component Erlang function to the spectra. The values of free parameters, normalization constant, and $\chi^2$/dof, and p-values are summarized in Table 3. We can see that the experimental data for all hadrons and energies are well described by the fit function. The values of $m_{2}$ are 2, and the values of $m_{1}$ are 2, 3, and 4. The values of weight factor $k_1$ are more than 50\%, and $N_0$ increases with increase of collision energy.

\begin{figure}[H]
\hskip-0.0cm {\centering
\includegraphics[width=16.0cm]{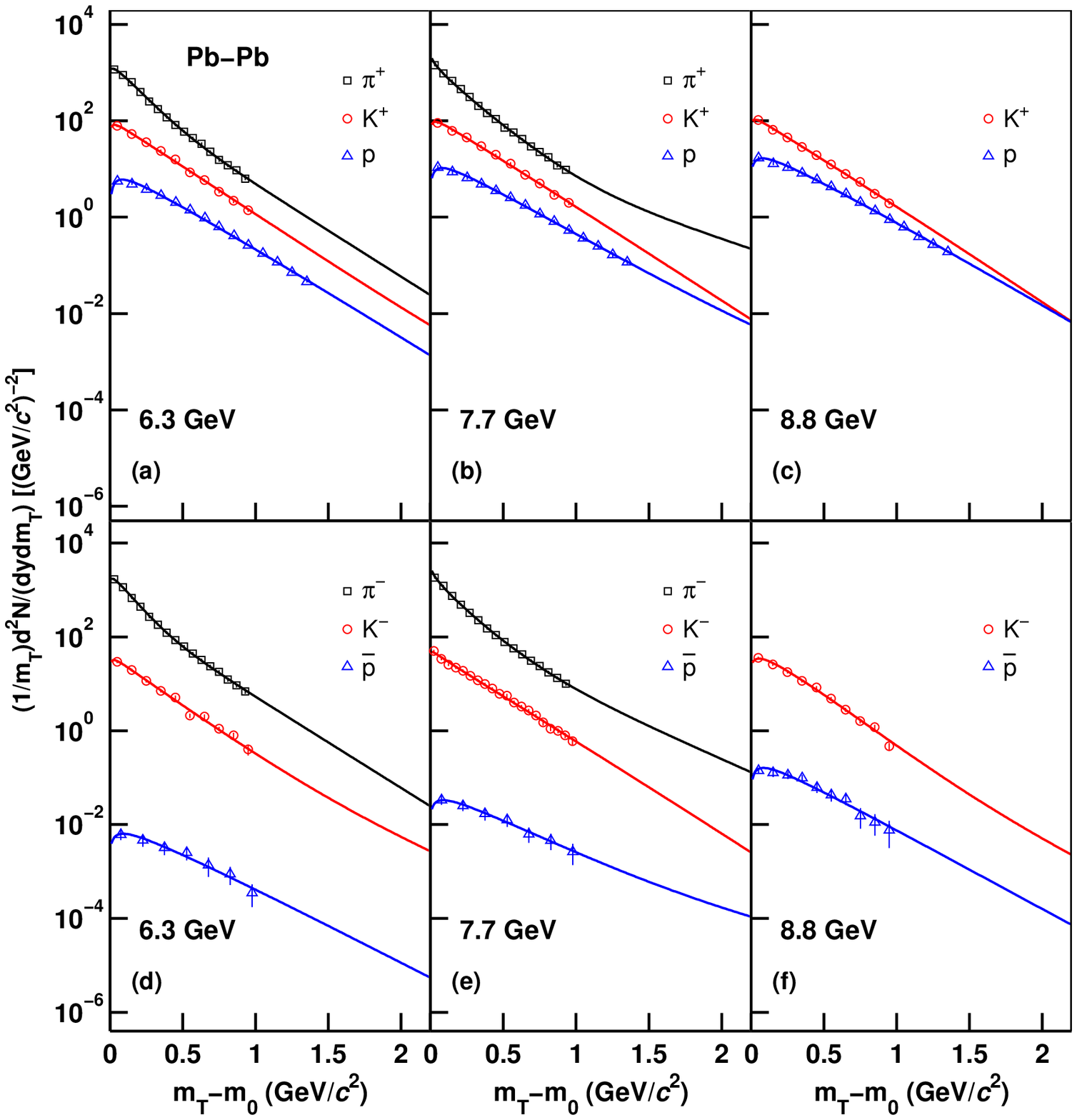}}
\vskip-0.18cm  Figure 5. Transverse mass spectra for $\pi^\pm$, $K^\pm$, $p$, and $\bar{p}$ in central Pb-Pb collisions at 6.3 [Figures 5(a) and 5(d)], 7.7 [Figures 5(b) and 5(e)], and 8.8 [Figures 5(c) and 5(f)] GeV . The symbols represent the experimental data taken by the NA49 Collaboration \cite{Alt(NA49Collaboration)2008,Alt(NA49Collaboration)2006,Afanasiev(NA49Collaboration)2002}. The errors are statistical only. The curves are fits of two-component Erlang function to the spectra.
\end{figure}

Figure 6 presents the $m_{T}$ and $p_{T}$ spectra of $\pi^\pm$, $K^\pm$, $p$, and $\bar{p}$ in central Pb-Pb collisions at 12.3 ([Figures 6(a) and 6(d)], 17.3 [Figures 6(b) and 6(e)], and 2760 [Figures 6(c) and 6(f)] GeV, where $\sigma_{trig}$ on the vertical axis denotes the interaction cross section satisfying a T0 centrality trigger. The symbols represent the experimental data reported by the NA49 Collaboration for 12.3 GeV at mid-rapidity [$|y|<0.1$ for $K^\pm$ \cite{Afanasiev(NA49Collaboration)2002}, $2.2 < y < 2.6$ ($y_{c.m.} = 2.57$) for $p$ and $\bar{p}$ \cite{Alt(NA49Collaboration)2006}], the NA44 Collaboration for 17.3 GeV near mid-rapidity ($2.4 < y < 3.1$ for $\pi^\pm$, $2.4 < y < 3.5$ for $K^\pm$, and $2.3 < y < 2.9$ for $p$ and $\bar{p}$ ) \cite{Bearden(NA44Collaboration)2002}, and the ALICE Collaboration for 2760 GeV at mid-rapidity $|y|<0.5$ \cite{Abelev(ALICECollaboration)2013}. Some data for different particles are divided by suitable factors for clarity. The errors are statistical for 12.3 GeV, are systematic for 17.3 GeV, and are quadratic sum of statistical errors and systematic errors for 2760 GeV. The curves represent the two-component Erlang fits. The values of free parameters, normalization constant, and $\chi^2$/dof, and p-values are summarized in Table 3. Obviously, the experimental data for all particles at all energies are in good agreement with the fits. The values of $m_{2}$ are 2, and the values of $m_{1}$ are 2, 3, and 4. The values of weight factor $k_1$ are more than 50\%, and $N_0$ increases with increase of collision energy.

Figure 7 shows the $p_{T}$ spectra of $\pi^\pm$ [Figures 7(a) and 7(b)] and $K^\pm$ [Figures 7(c) and 7(d)] produced in mid-rapidity $y \approx 0$ inelastic $pp$ collisions at 6.3, 7.7, 8.8, 12.3, and 17.3 GeV. The measurements were performed at the CERN-Super Proton Synchrotron (SPS) by the large acceptance NA61/SHINE hadron spectrometer \cite{Aduszkiewicz(NA61/SHINECollaboration)2017}. Spectra at different energies are scaled by appropriate factors for better visibility. The error bars on data points correspond to combined statistical and systematic uncertainties. The curves are our fitting results by using the one- or two-component Erlang function. For some curves, we use one-component Erlang function because the number of corresponding experimental data points is small. Due to the proportion of the second component is small, it has little effect on the calculated particle ratio, despite the absence of the second component. The values of free parameters, normalization constant, and $\chi^2$/dof, and p-values are given in Table 4. As can be seen, the fits for all hadrons at all energies are in good agreement with the experimental data. The values of $m_{2}$ and $m_{1}$ are 2 and 3, respectively. The values of weight factor $k_1$ are more than 50\%. It should be noted that the dof for $\pi^-$ at 12.3 GeV in Table 4 is zero, which means the dash curve in Figure 7(b) is drawn to guide the eye.

Figure 8 presents the $p_{T}$ spectra of $\pi^\pm$, $K^\pm$, $p$, and $\bar{p}$ produced in inelastic $pp$ collisions at 62.4 [Figures 8(a) and 8(c)] and 200 [Figures 8(b) and 8(d)] GeV, where $E$ and $\sigma$ on the vertical axis denote the particle energy and cross-section, respectively. The data measured by the PHENIX Collaboration in the mid-pseudorapidity range $|\eta|<0.35$\cite{Adare2011}, are represented in different panels by different symbols. Spectra for different particles are scaled by appropriate factors for better visibility. The error bars are statistical only. The curves are our results fitted by using the two-component Erlang distribution. The values of free parameters, normalization constant, and $\chi^2$/dof, and p-values are given in Table 4. One can see that all the fitting results by using the two-component Erlang function are consistent with the experimental data. The values of $m_{2}$ are 2, and the values of $m_{1}$ are 2, 3, and 4. The values of weight factor $k_1$ are more than 50\%, and $N_0$ increases with increase of collision energy.

Figure 9 exhibits the $p_{T}$ spectra of $\pi^\pm$, $K^\pm$, $p$, and $\bar{p}$ produced in inelastic $pp$ collisions at 900 GeV [Figures 9(a) and 9(c)] and 2.76 TeV [Figures 9(b) and 9(d)]. The symbols also denote the experimental data recorded by the CMS Collaboration in the range $|y|<1$\cite{Chatrchyan2012}. The data for different particles are scaled by suitable factors for clarity. The error bars indicate the combined uncorrelated statistical and systematic uncertainties, and the fully correlated normalization uncertainty is 3.0\%. The curves are fits of the two-component Erlang function to the spectra. The values of free parameters, normalization constant, and $\chi^2$/dof, and p-values are summarized in Table 5. We can see that the experimental data are well described by the fit function. The values of $m_{2}$ are 2, and the values of $m_{1}$ are 2, 3, and 4. The values of weight factor $k_1$ are more than 50\%, and $N_0$ increases with increase of collision energy.

\begin{figure}[H]
\hskip-0.0cm {\centering
\includegraphics[width=16.0cm]{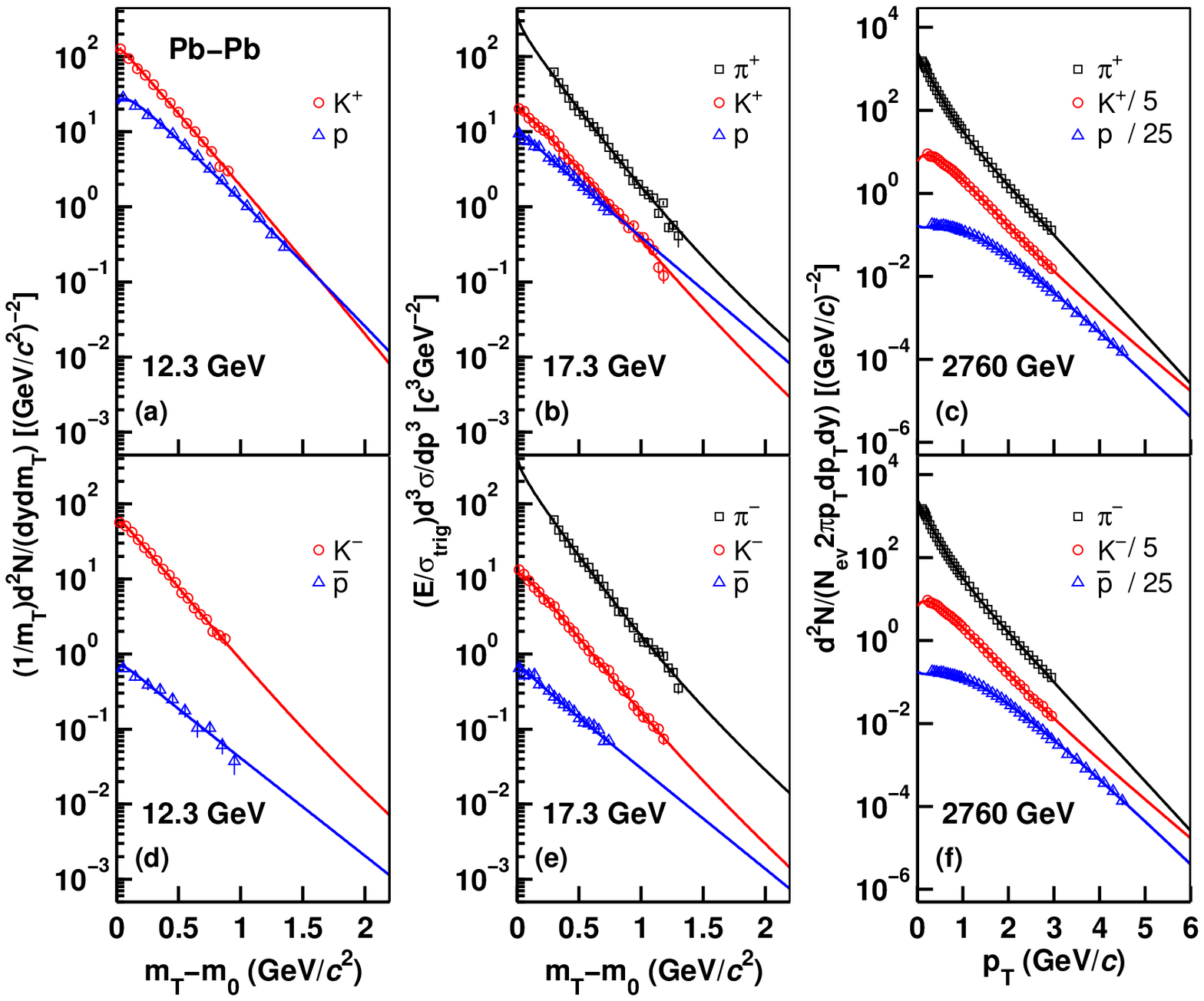}}
\vskip-0.18cm  Figure 6. Transverse mass and momentum spectra for $\pi^\pm$, $K^\pm$, $p$, and $\bar{p}$ at mid-rapidity in central Pb-Pb collisions at 12.3 [Figures 6(a) and 6(d)], 17.3 [Figures 6(b) and 6(e)], and 2760 [Figures 6(c) and 6(f)] GeV. The symbols represent the experimental data reported by the NA49 Collaboration for 12.3 GeV at mid-rapidity [$|y|<0.1$ for $K^\pm$ \cite{Afanasiev(NA49Collaboration)2002}, $2.2 < y < 2.6$ ($y_{c.m.} = 2.57$) for $p$ and $\bar{p}$ \cite{Alt(NA49Collaboration)2006}], the NA44 Collaboration for 17.3 GeV near mid-rapidity ($2.4 < y < 3.1$ for $\pi^\pm$, $2.4 < y < 3.5$ for $K^\pm$, and $2.3 < y < 2.9$ for $p$ and $\bar{p}$ ) \cite{Bearden(NA44Collaboration)2002}, and the ALICE Collaboration for 2760 GeV at mid-rapidity $|y|<0.5$ \cite{Abelev(ALICECollaboration)2013}. Some data for different particles are divided by suitable factors for clarity. The errors are statistical for 12.3 GeV, are systematic for 17.3 GeV, and are quadratic sum of statistical errors and systematic errors for 2760 GeV. The curves are fits of two-component Erlang function to the spectra.
\end{figure}

\newpage
{\scriptsize {Table 3. Values of free parameters, normalization constant, and $\chi^2$/dof, and p-values corresponding to two-component Erlang $p_T$ (or $m_T$) distribution for Pb-Pb collisions in Figures 5 and 6.
{%
\begin{center}
\begin{tabular}{ccccccccccc}
\hline
\hline
Figure  & $\sqrt{s_{NN}}$ & Particle & $m_1$ &$<p_{ti1}>$ & $k_1$ & $m_2$ &$<p_{ti2}>$ & $N_0$& $\chi^2$/dof & p-value\\
        &  (GeV)          &          &       & (GeV/c)    &       &       & (GeV/c)    &      &              &        \\
\hline
Figure 5 (a) & 6.3  & $\pi^{+}$      & 3 & 0.095 $\pm$ 0.003 & 0.51 $\pm$ 0.03 & 2 & 0.228 $\pm$ 0.005 & 72.088 $\pm$ 2.379 & 9.764/10 & 0.461\\
             &      & $K^{+}$        & 3 & 0.194 $\pm$ 0.004 & 0.90 $\pm$ 0.10 & 2 & 0.295 $\pm$ 0.059 & 16.508 $\pm$ 0.644 & 6.875/4 & 0.143\\
             &      & $p$            & 4 & 0.209 $\pm$ 0.002 & 1               & - & -                 &  2.889 $\pm$ 0.078 & 27.988/11 & 0.003\\
Figure 5 (d) &      & $\pi^{-}$      & 3 & 0.082 $\pm$ 0.003 & 0.51 $\pm$ 0.03 & 2 & 0.225 $\pm$ 0.006 & 83.773 $\pm$ 2.765 & 4.884/10 & 0.899\\
             &      & $K^{-}$        & 3 & 0.179 $\pm$ 0.005 & 0.86 $\pm$ 0.09 & 2 & 0.328 $\pm$ 0.081 &  5.644 $\pm$ 0.186 & 13.617/4 & 0.009\\
             &      & $\overline{p}$ & 4 & 0.234 $\pm$ 0.018 & 1               & - & -                 &  0.004 $\pm$ 0.001 & 1.420/4 & 0.841\\
\hline
Figure 5 (b) & 7.7  & $\pi^{+}$      & 2 & 0.172 $\pm$ 0.003 & 0.90 $\pm$ 0.02 & 2 & 0.436 $\pm$ 0.079 & 87.611 $\pm$ 2.541 & 10.081/10 & 0.433\\
             &      & $K^{+}$        & 3 & 0.202 $\pm$ 0.004 & 1               & - & -                 & 20.584 $\pm$ 0.638 & 6.746/4 & 0.456\\
             &      & $p$            & 4 & 0.214 $\pm$ 0.004 & 0.87 $\pm$ 0.12 & 2 & 0.420 $\pm$ 0.055 &  5.278 $\pm$ 0.306 & 2.421/8 & 0.965\\
Figure 5 (e) &      & $\pi^{-}$      & 2 & 0.151 $\pm$ 0.004 & 0.79 $\pm$ 0.03 & 2 & 0.317 $\pm$ 0.022 & 98.711 $\pm$ 3.257 & 4.493/10 & 0.922\\
             &      & $K^{-}$        & 3 & 0.205 $\pm$ 0.006 & 0.82 $\pm$ 0.08 & 2 & 0.220 $\pm$ 0.038 &  8.170 $\pm$ 0.302 & 18.012/14 & 0.206\\
             &      & $\overline{p}$ & 4 & 0.233 $\pm$ 0.028 & 0.75 $\pm$ 0.15 & 2 & 0.580 $\pm$ 0.116 &  0.021 $\pm$ 0.004 & 0.4253/1 & 0.514\\
\hline
Figure 5 (c) & 8.8  & $K^{+}$        & 3 & 0.203 $\pm$ 0.004 & 0.91 $\pm$ 0.05 & 2 & 0.205 $\pm$ 0.041 & 21.556 $\pm$ 0.625 & 5.397/4 & 0.249\\
             &      & $p$            & 4 & 0.221 $\pm$ 0.003 & 0.92 $\pm$ 0.08 & 2 & 0.270 $\pm$ 0.054 &  8.605 $\pm$ 0.422 & 7.546/8 & 0.479\\
Figure 5 (f) &      & $K^{-}$        & 4 & 0.149 $\pm$ 0.004 & 0.75 $\pm$ 0.15 & 2 & 0.280 $\pm$ 0.031 &  8.062 $\pm$ 0.282 & 13.337/4 & 0.010\\
             &      & $\overline{p}$ & 4 & 0.219 $\pm$ 0.010 & 0.92 $\pm$ 0.07 & 2 & 0.370 $\pm$ 0.074 &  0.084 $\pm$ 0.010 & 3.408/4 & 0.492\\
\hline
Figure 6 (a) & 12.3 & $K^{+}$        & 3 & 0.204 $\pm$ 0.004 & 0.90 $\pm$ 0.07 & 2 & 0.220 $\pm$ 0.027 & 24.872 $\pm$ 0.696 & 6.261/8 & 0.618\\
             &      & $p$            & 4 & 0.224 $\pm$ 0.003 & 0.85 $\pm$ 0.07 & 2 & 0.290 $\pm$ 0.046 & 13.871 $\pm$ 0.569 & 8.960/8 & 0.346\\
Figure 6 (d) &      & $K^{-}$        & 3 & 0.192 $\pm$ 0.004 & 0.83 $\pm$ 0.07 & 2 & 0.327 $\pm$ 0.065 & 11.307 $\pm$ 0.339 & 5.198/12 & 0.951\\
             &      & $\overline{p}$ & 3 & 0.312 $\pm$ 0.010 & 0.94 $\pm$ 0.06 & 2 & 0.370 $\pm$ 0.074 &  0.348 $\pm$ 0.022 & 3.947/4 & 0.413\\
\hline
Figure 6 (b) & 17.3 & $\pi^{+}$      & 2 & 0.199 $\pm$ 0.004 & 0.91 $\pm$ 0.04 & 2 & 0.330 $\pm$ 0.038 & 17.753 $\pm$ 0.781 & 44.048/20 & 0.001\\
             &      & $K^{+}$        & 4 & 0.162 $\pm$ 0.005 & 0.51 $\pm$ 0.06 & 2 & 0.289 $\pm$ 0.014 &  4.246 $\pm$ 0.238 & 19.806/24 & 0.708\\
             &      & $p$            & 3 & 0.292 $\pm$ 0.008 & 0.96 $\pm$ 0.02 & 2 & 0.322 $\pm$ 0.020 &  0.403 $\pm$ 0.014 & 6.545/12 & 0.886\\
Figure 6 (e) &      & $\pi^{-}$      & 2 & 0.188 $\pm$ 0.003 & 0.88 $\pm$ 0.03 & 2 & 0.307 $\pm$ 0.014 & 18.926 $\pm$ 0.662 & 24.600/20 & 0.217\\
             &      & $K^{-}$        & 4 & 0.150 $\pm$ 0.005 & 0.51 $\pm$ 0.10 & 2 & 0.286 $\pm$ 0.011 &  2.378 $\pm$ 0.131 & 9.820/24 & 0.995\\
             &      & $\overline{p}$ & 3 & 0.304 $\pm$ 0.011 & 0.93 $\pm$ 0.03 & 2 & 0.170 $\pm$ 0.034 &  0.028 $\pm$ 0.002 & 7.744/12 & 0.805\\
\hline
Figure 6 (c) & 2760 & $\pi^{+}$      & 2 & 0.177 $\pm$ 0.007 & 0.58 $\pm$ 0.03 & 2 & 0.364 $\pm$ 0.004 &118.422 $\pm$ 3.790 & 14.385/35 & 0.999\\
             &      & $K^{+}$        & 3 & 0.283 $\pm$ 0.009 & 0.64 $\pm$ 0.10 & 2 & 0.473 $\pm$ 0.021 & 17.258 $\pm$ 0.552 & 2.657/30 & 1\\
             &      & $p$            & 4 & 0.360 $\pm$ 0.005 & 0.82 $\pm$ 0.05 & 2 & 0.470 $\pm$ 0.064 &  5.311 $\pm$ 0.175 & 9.093/36 & 1\\
Figure 6 (f) &      & $\pi^{-}$      & 2 & 0.180 $\pm$ 0.007 & 0.58 $\pm$ 0.03 & 2 & 0.364 $\pm$ 0.004 &118.191 $\pm$ 3.782 & 12.640/35 & 0.999\\
             &      & $K^{-}$        & 3 & 0.277 $\pm$ 0.010 & 0.59 $\pm$ 0.09 & 2 & 0.467 $\pm$ 0.016 & 17.198 $\pm$ 0.602 & 2.851/30 & 1\\
             &      & $\overline{p}$ & 4 & 0.360 $\pm$ 0.006 & 0.81 $\pm$ 0.07 & 2 & 0.470 $\pm$ 0.076 &  5.253 $\pm$ 0.173 & 10.028/36 & 1\\
\hline
\end{tabular}
\end{center}
}} }

\begin{figure}[H]
\hskip-0.0cm {\centering
\includegraphics[width=16.0cm]{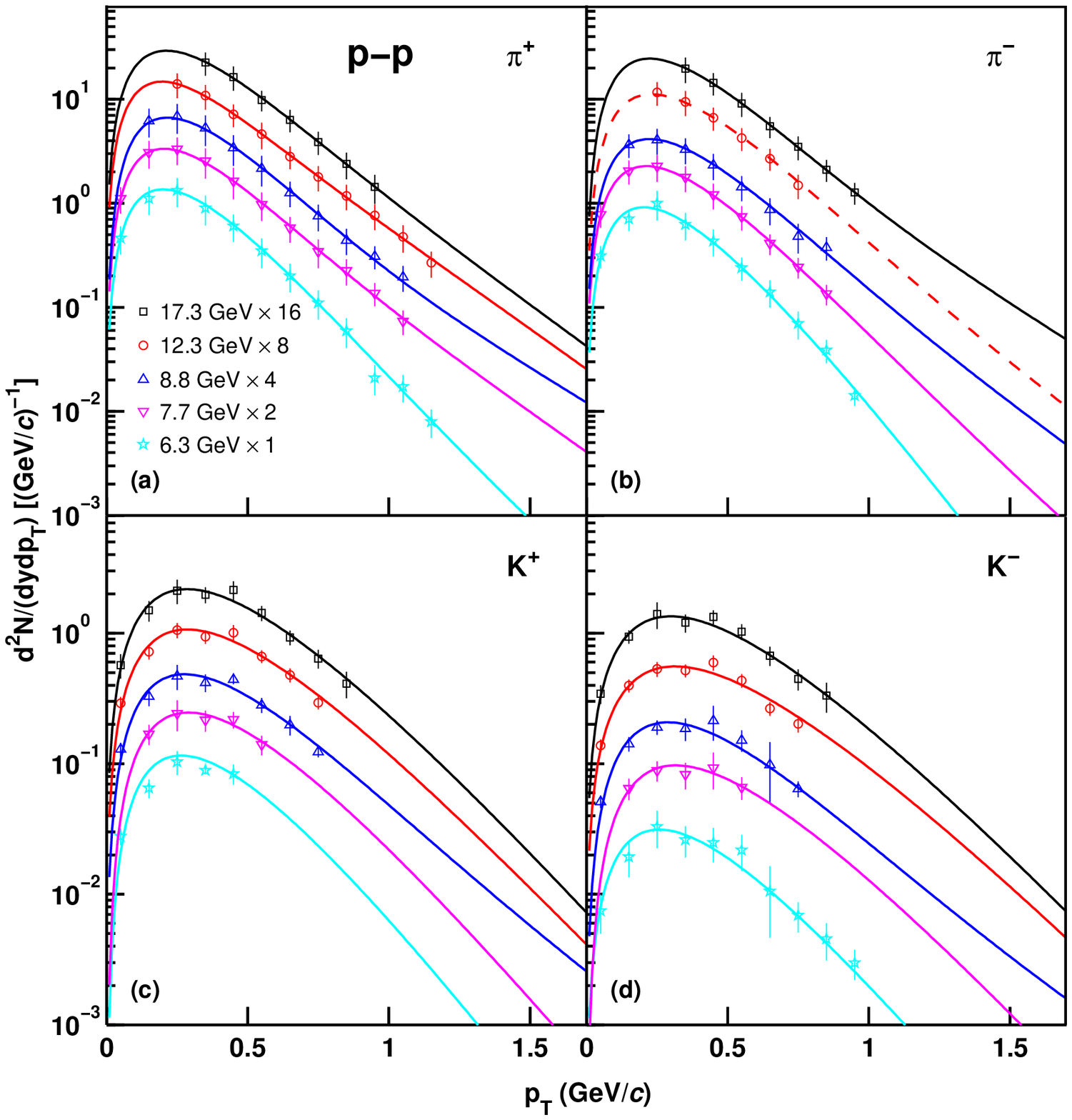}}
\vskip-0.18cm  Figure 7. Transverse momentum spectra for$\pi^\pm$ [Figures 7(a) and 7(b)] and $K^\pm$ [Figures 7(c) and 7(d)] in $y \approx 0$ inelastic $pp$ collisions at SPS energies (6.3, 7.7, 8.8, 12.3, and 17.3 GeV). The symbols represent the experimental data reported by the NA61/SHINE Collaboration \cite{Aduszkiewicz(NA61/SHINECollaboration)2017}. Spectra at different energies are scaled by appropriate factors for better visibility. The errors are quadratic sum of statistical errors and systematic errors. The curves are fits of one- or two-component Erlang function to the spectra, where the dash curve in the figure is drawn to guide the eye.
\end{figure}

\begin{figure}[H]
\hskip-0.0cm {\centering
\includegraphics[width=16.0cm]{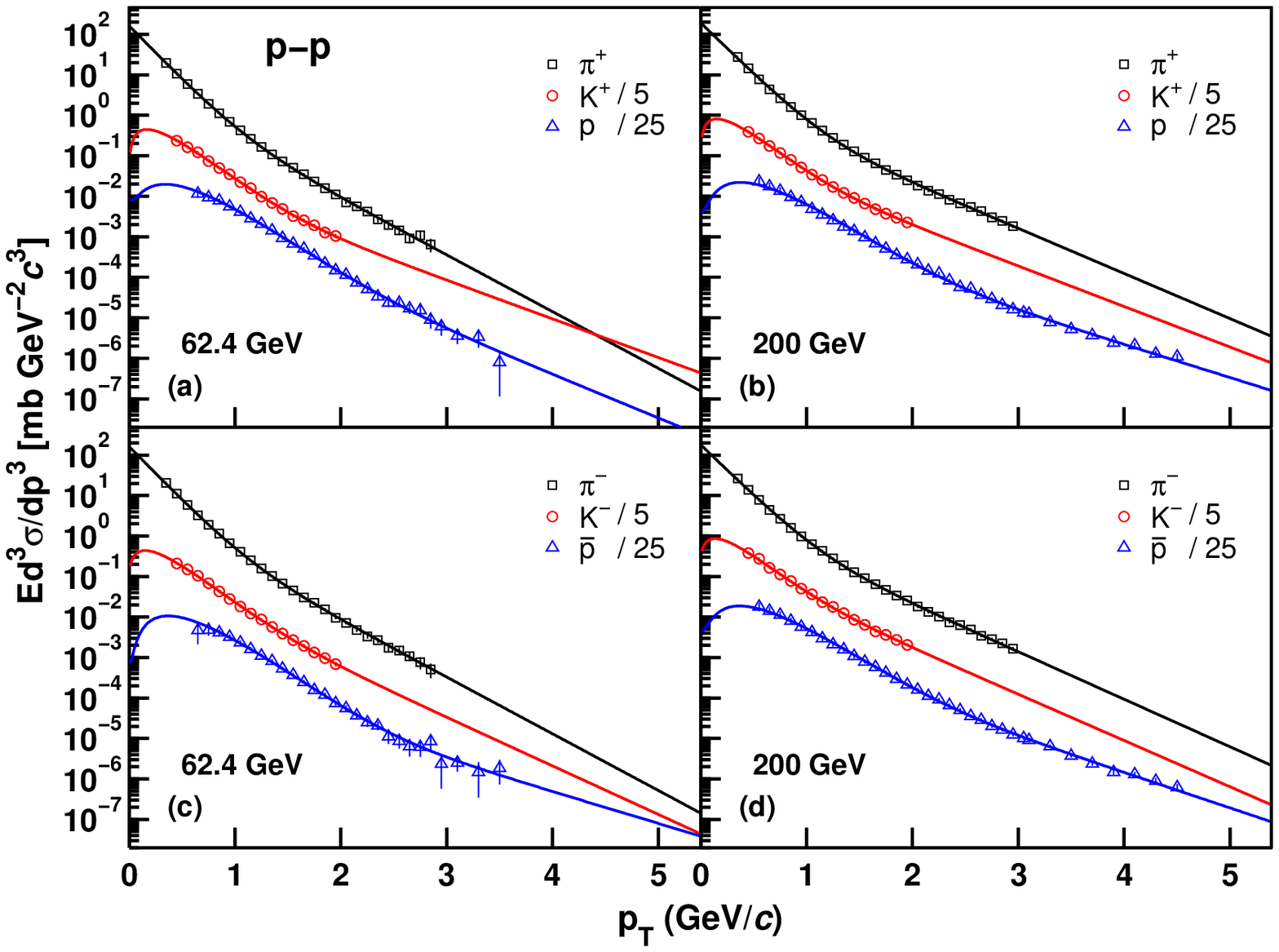}}
\vskip-0.18cm Figure 8. Transverse momentum spectra for positive ($\pi^+$, $K^+$, $p$) and negative ($\pi^-$, $K^-$, $\bar{p}$) particles produced in inelastic $pp$ collisions at 62.4 [Figures 8(a) and 8(c)] and 200 [Figures 8(b) and 8(d)] GeV. The experimental data represented by the symbols are measured by the PHENIX Collaboration in $|\eta|<0.35$\cite{Adare2011}. The errors, are statistical only. The curves are fits by the two-component Erlang distribution.
\end{figure}

\newpage
{\scriptsize {Table 4. Values of free parameters, normalization constant, and $\chi^2$/dof, and p-values corresponding to one- or two-component Erlang $p_T$ distribution for inelastic $pp$ collisions in Figures 7 and 8.
{%
\begin{center}
\begin{tabular}{ccccccccccc}
\hline
\hline
Figure  & $\sqrt{s_{NN}}$ & Particle & $m_1$ &$<p_{ti1}>$ & $k_1$ & $m_2$ &$<p_{ti2}>$ & $N_0$& $\chi^2$/dof & p-value\\
        &  (GeV)          &          &       & (GeV/c)    &       &       & (GeV/c)    &      & &              \\
\hline
Figure 7 (a) & 6.3 & $\pi^{+}$ & 3 & 0.103 $\pm$ 0.006 & 0.77 $\pm$ 0.09 & 2 & 0.163 $\pm$ 0.011 & 0.538 $\pm$ 0.051 & 2.193/6 & 0.901\\
Figure 7 (c) &     & $K^{+}$   & 3 & 0.143 $\pm$ 0.012 & 1               & - & -                 & 0.056 $\pm$ 0.001 & 12.005/2 & 0.002\\
Figure 7 (b) &     & $\pi^{-}$ & 3 & 0.106 $\pm$ 0.002 & 0.90 $\pm$ 0.06 & 2 & 0.116 $\pm$ 0.010 & 0.358 $\pm$ 0.038 & 1.874/4 & 0.759\\
Figure 7 (d) &     & $K^{-}$   & 3 & 0.135 $\pm$ 0.008 & 0.80 $\pm$ 0.07 & 2 & 0.184 $\pm$ 0.020 & 0.016 $\pm$ 0.003 & 2.566/4 & 0.632\\
\hline
Figure 7 (a) & 7.7 & $\pi^{+}$ & 3 & 0.102 $\pm$ 0.008 & 0.69 $\pm$ 0.08 & 2 & 0.202 $\pm$ 0.011 & 0.698 $\pm$ 0.086 & 0.246/5 & 0.998\\
Figure 7 (c) &     & $K^{+}$   & 3 & 0.145 $\pm$ 0.013 & 1               & - & -                 & 0.066 $\pm$ 0.006 & 0.862/2 & 0.650\\
Figure 7 (b) &     & $\pi^{-}$ & 3 & 0.110 $\pm$ 0.042 & 0.73 $\pm$ 0.12 & 2 & 0.170 $\pm$ 0.012 & 0.482 $\pm$ 0.078 & 0.084/3 & 0.993\\
Figure 7 (d) &     & $K^{-}$   & 3 & 0.158 $\pm$ 0.017 & 1               & - & -                 & 0.028 $\pm$ 0.001 & 0.487/2 & 0.784\\
\hline
Figure 7 (a) & 8.8 & $\pi^{+}$ & 3 & 0.107 $\pm$ 0.004 & 0.80 $\pm$ 0.04 & 2 & 0.232 $\pm$ 0.015 & 0.711 $\pm$ 0.055 & 0.370/4 & 0.985\\
Figure 7 (c) &     & $K^{+}$   & 3 & 0.144 $\pm$ 0.004 & 0.75 $\pm$ 0.07 & 2 & 0.190 $\pm$ 0.060 & 0.067 $\pm$ 0.001 & 5.585/2 & 0.061\\
Figure 7 (b) &     & $\pi^{-}$ & 3 & 0.113 $\pm$ 0.005 & 0.75 $\pm$ 0.06 & 2 & 0.203 $\pm$ 0.018 & 0.461 $\pm$ 0.038 & 0.653/2 & 0.722\\
Figure 7 (d) &     & $K^{-}$   & 3 & 0.162 $\pm$ 0.007 & 0.75 $\pm$ 0.08 & 2 & 0.193 $\pm$ 0.090 & 0.030 $\pm$ 0.001 & 1.179/2 & 0.411\\
\hline
Figure 7 (a) & 12.3& $\pi^{+}$ & 3 & 0.098 $\pm$ 0.009 & 0.51 $\pm$ 0.04 & 2 & 0.198 $\pm$ 0.008 & 0.796 $\pm$ 0.068 & 0.212/4 & 0.995\\
Figure 7 (c) &     & $K^{+}$   & 3 & 0.155 $\pm$ 0.005 & 0.76 $\pm$ 0.07 & 2 & 0.180 $\pm$ 0.030 & 0.076 $\pm$ 0.003 & 4.164/2 & 0.125\\
Figure 7 (b) &     &$\pi^{-}$  & 3 & 0.117 $\pm$ 0.009 & 0.80 $\pm$ 0.09 & 2 & 0.202 $\pm$ 0.016 & 0.631 $\pm$ 0.089 & 0.165/0 & 0\\
Figure 7 (d) &     & $K^{-}$   & 3 & 0.159 $\pm$ 0.004 & 0.64 $\pm$ 0.06 & 2 & 0.246 $\pm$ 0.030 & 0.044 $\pm$ 0.001 & 4.512/2 & 0.105\\
\hline
Figure 7 (a) & 17.3& $\pi^{+}$ & 3 & 0.108 $\pm$ 0.008 & 0.61 $\pm$ 0.08 & 2 & 0.198 $\pm$ 0.017 & 0.813 $\pm$ 0.049 & 0.054/1 & 0.817\\
Figure 7 (c) &     & $K^{+}$   & 3 & 0.154 $\pm$ 0.007 & 0.80 $\pm$ 0.06 & 2 & 0.178 $\pm$ 0.030 & 0.077 $\pm$ 0.006 & 2.300/3 & 0.512\\
Figure 7 (b) &     & $\pi^{-}$ & 3 & 0.112 $\pm$ 0.006 & 0.77 $\pm$ 0.06 & 2 & 0.227 $\pm$ 0.020 & 0.688 $\pm$ 0.004 & 0.032/1 & 0.859\\
Figure 7 (d) &     & $K^{-}$   & 3 & 0.164 $\pm$ 0.008 & 0.80 $\pm$ 0.07 & 2 & 0.178 $\pm$ 0.030 & 0.051 $\pm$ 0.003 & 3.057/3 & 0.382\\
\hline
Figure 8 (a) & 62.4& $\pi^{+}$ & 2 & 0.160 $\pm$ 0.003 & 0.88 $\pm$ 0.01 & 2 & 0.310 $\pm$ 0.005 & 0.965 $\pm$ 0.039 & 5.886/26 & 1\\
             &     & $K^{+}$   & 3 & 0.175 $\pm$ 0.003 & 0.84 $\pm$ 0.02 & 2 & 0.455 $\pm$ 0.019 & 0.085 $\pm$ 0.003 & 3.026/16 & 1\\
             &     & $p$       & 4 & 0.186 $\pm$ 0.004 & 0.83 $\pm$ 0.04 & 2 & 0.404 $\pm$ 0.016 & 0.040 $\pm$ 0.002 & 6.489/27 & 1\\
Figure 8 (c)  &     & $\pi^{-}$ & 2 & 0.160 $\pm$ 0.003 & 0.89 $\pm$ 0.01 & 2 & 0.310 $\pm$ 0.004 & 0.966 $\pm$ 0.035 & 8.208/26 & 1\\
             &     & $K^{-}$   & 3 & 0.168 $\pm$ 0.004 & 0.76 $\pm$ 0.03 & 2 & 0.362 $\pm$ 0.014 & 0.076 $\pm$ 0.003 & 1.856/16 & 1\\
             &     & $\overline{p}$ & 4 & 0.186 $\pm$ 0.003 & 0.95 $\pm$ 0.02 & 2 & 0.552 $\pm$ 0.046 & 0.022 $\pm$ 0.001 & 7.113/27 & 1\\
\hline
Figure 8 (b)  & 200 & $\pi^{+}$ & 2 & 0.170 $\pm$ 0.003 & 0.91 $\pm$ 0.01 & 2 & 0.390 $\pm$ 0.004 & 1.071 $\pm$ 0.041 & 11.062/27 & 0.997\\
             &     & $K^{+}$   & 3 & 0.162 $\pm$ 0.004 & 0.73 $\pm$ 0.02 & 2 & 0.434 $\pm$ 0.012 & 0.120 $\pm$ 0.005 & 1.123/16 & 1\\
             &     & $p$       & 4 & 0.194 $\pm$ 0.004 & 0.88 $\pm$ 0.01 & 2 & 0.532 $\pm$ 0.008 & 0.043 $\pm$ 0.002 & 17.121/34 & 0.993\\
Figure 8 (d)  &     & $\pi^{-}$ & 2 & 0.170 $\pm$ 0.003 & 0.89 $\pm$ 0.01 & 2 & 0.370 $\pm$ 0.004 & 1.028 $\pm$ 0.040 & 8.247/27 & 1\\
             &     & $K^{-}$   & 3 & 0.156 $\pm$ 0.005 & 0.64 $\pm$ 0.03 & 2 & 0.380 $\pm$ 0.009 & 0.119 $\pm$ 0.005 & 3.264/16 & 1\\
             &     & $\overline{p}$ & 4 & 0.192 $\pm$ 0.004 & 0.86 $\pm$ 0.02 & 2 & 0.497 $\pm$ 0.006 & 0.036 $\pm$ 0.002 & 10.592/34 & 1\\
\hline
\end{tabular}
\end{center}
}} }

Figure 10 shows the same as Fig. 9, but for $\sqrt{s}=$ 7 TeV [Figures 10(a) and 10(c)] and $\sqrt{s}=$ 13 TeV [Figures 10(b) and 10(d)]. The symbols also denote the experimental data recorded by the CMS collaboration in the range $|y|<1$\cite{Chatrchyan2012,Sirunyan(CMSCollaboration)2017}. The error bars indicate the combined uncorrelated statistical and systematic uncertainties, and the fully correlated normalization uncertainty is 3.0\%. The curves are our results fitted by using the two-component Erlang distribution. The values of free parameters, normalization constant, and $\chi^2$/dof, and p-values are summarized in Table 5. It is not hard to see that the experimental data can be well fitted by the two-component Erlang distribution. Similarly, The values of $m_{2}$ are 2, and the values of $m_{1}$ are 2 and 3. The values of weight factor $k_1$ are more than 50\%, and $N_0$ increases with increase of collision energy.

According to the extracted normalization constants from the above comparisons, the three types of energy dependent yield ratios, $\pi^{-}/\pi^{+}$, $K^{-}/K^{+}$, and $\overline{p}/p$, of negative to positive particles from different collision systems, are obtained and are shown in Figure 11. The black, red, and blue circles denote respectively the calculated results from inelastic $pp$, central Au-Au, and central Pb-Pb collisions. For comparison, the black, red, and blue triangles correspondingly denote the experimental results\cite{Adamczyk(STARCollaboration)2017,Klay(E895Collaboration)2003,Ahle(E-802Collaboration)1998,Ahle(E866andE917Collaborations)2000,Abelev(STARCollaboration)2009paper2,
Adcox(PHENIXCollaboration)2004,Adler(PHENIXCollaboration)2004,Alt(NA49Collaboration)2008,Alt(NA49Collaboration)2006,Afanasiev(NA49Collaboration)2002,Bearden(NA44Collaboration)2002,
Abelev(ALICECollaboration)2013,Aduszkiewicz(NA61/SHINECollaboration)2017,Sirunyan(CMSCollaboration)2017,Chatrchyan2012,Klay(E895Collaboration)2002,Akiba(E802Collaboration)1996,
Adcox(PHENIXCollaboration)2002,Abelev(STARCollaboration)2010,Adams(STARCollaboration)2004,Afanasiev(NA49Collaboration)2004,Abelev(STARCollaboration)2007,Aamodt(ALICECollaboration)2011} from inelastic or NSD $pp$, central Au-Au, and central Pb-Pb collisions, respectively. One can see that our calculation results based on transverse momentum (or mass) spectra are consistent with the experimental data. That means the differences between our calculation results and experimental data are very small, which indicates that our calculation method is correct.

In fact, the yield ratios we calculate on the basis of transverse momentum (or mass) spectra, are at the stage of kinetic freeze-out and are affected by strong decay from high-mass resonance and weak decay from heavy flavor hadrons. In order to obtain the yield ratios at the stage of chemical freeze-out, one need remove the contributions of strong decay and weak decay from the above yield ratios. According to reference\cite{Yu2019}, we remove the contributions of strong and weak decays and obtain the modified (primary) yield ratios, $k_{\pi}$, $k_{K}$, and $k_{p}$. The calculation result shows that strong decay affects mainly $k_{\pi}$ and $k_{K}$, and weak decay affects mainly $k_{p}$. Strong decay can pull down $k_{\pi}$ and lift $k_{K}$, and weak decay can lift $k_{p}$, although these two decays do not significantly affect the primary yield ratios as a whole.

\begin{figure}[H]
\hskip-0.0cm {\centering
\includegraphics[width=16.0cm]{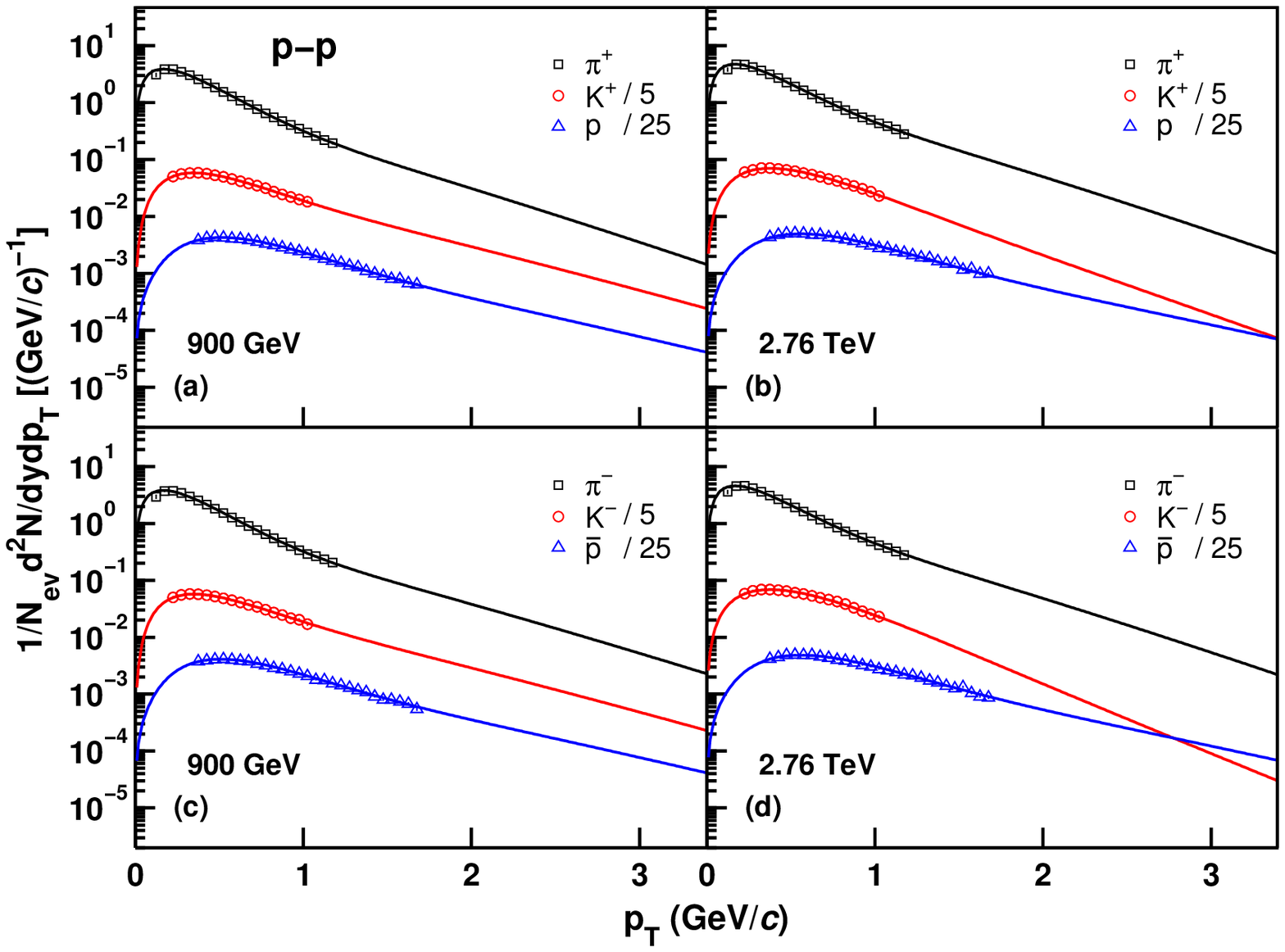}}
\vskip-0.18cm  Figure 9. Transverse momentum spectra for positive ($\pi^+$, $K^+$, $p$) and negative ($\pi^-$, $K^-$, $\bar{p}$) particles produced in inelastic $pp$ collisions at 900 GeV [Figures 9(a) and 9(c)] and 2.76 TeV [Figures 9(b) and 9(d)]. The symbols represent the experimental data recorded by the CMS Collaboration in $|y|<1$\cite{Chatrchyan2012}. The errors are the combined uncorrelated statistical and systematic ones, and the fully correlated normalization uncertainty is 3.0\%. The curves are our results fitted by the two-component Erlang distribution.
\end{figure}

\begin{figure}[H]
\hskip-0.0cm {\centering
\includegraphics[width=16.0cm]{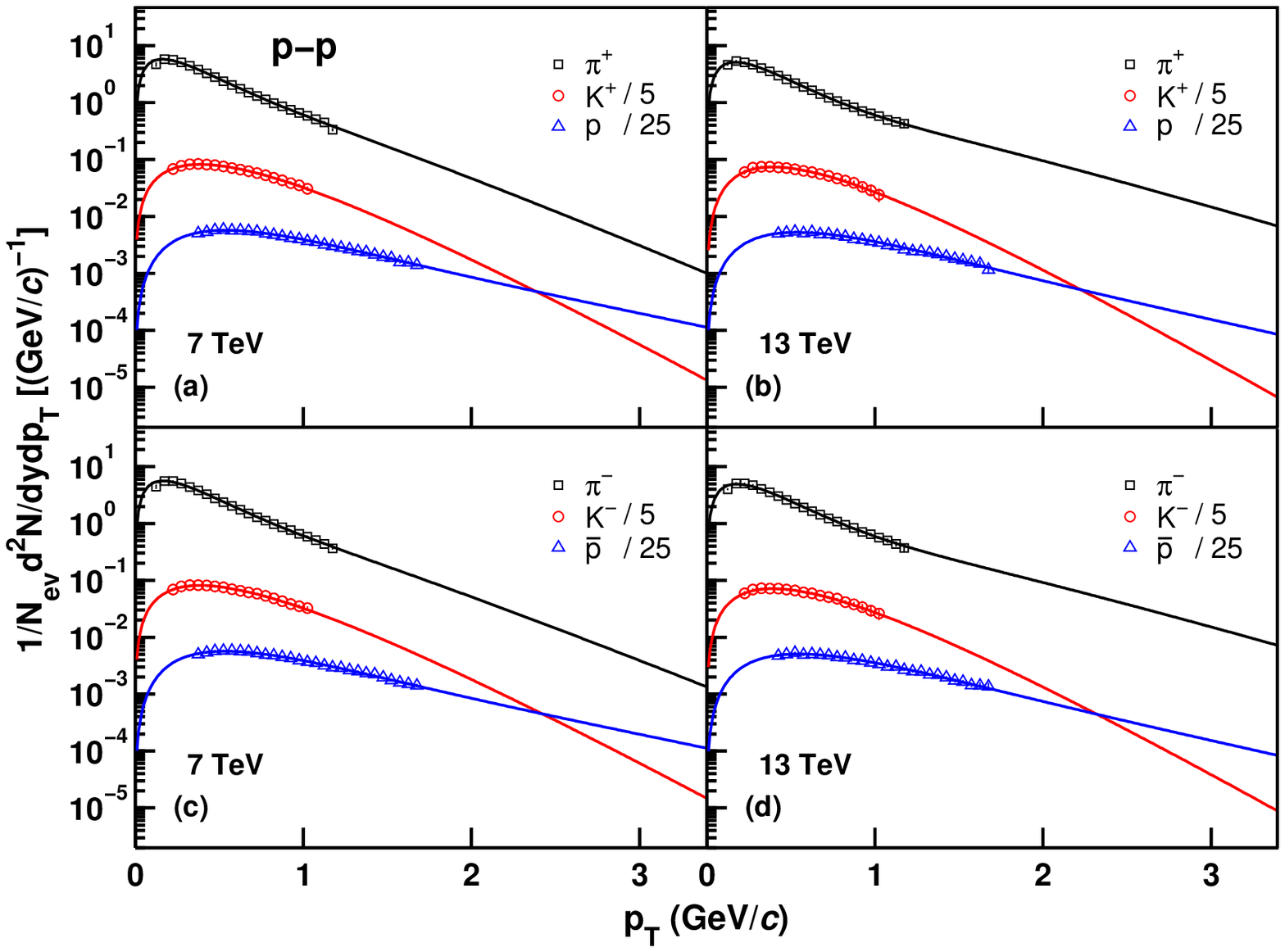}}
\vskip-0.18cm  Figure 10. The same $p_{T}$ spectra as Fig. 9,  but for $\sqrt{s}=$ 7 TeV [Figures 10(a) and 10(c)] and $\sqrt{s}=$ 13 TeV [Figures 10(b) and 10(d)]. The symbols also represent the experimental data recorded by the CMS Collaboration in $|y|<1$\cite{Chatrchyan2012,Sirunyan(CMSCollaboration)2017}. Similarly, the errors are the combined uncorrelated statistical and systematic ones, and the fully correlated normalization uncertainty is 3.0\%. The curves are our results fitted by the two-component Erlang distribution.
\end{figure}
\newpage
{\scriptsize {Table 5. Values of free parameters, normalization constant, and $\chi^2$/dof, and p-values corresponding to one- or two-component Erlang $p_T$ distribution for inelastic $pp$ collisions in Figures 9 and 10.
{%
\begin{center}
\begin{tabular}{ccccccccccc}
\hline
\hline
Figure  & $\sqrt{s_{NN}}$ & Particle & $m_1$ &$<p_{ti1}>$ & $k_1$ & $m_2$ &$<p_{ti2}>$ & $N_0$& $\chi^2$/dof & p-value\\
        &  (GeV)          &          &       & (GeV/c)    &       &       & (GeV/c)    &      &    &           \\
\hline
Figure 9 (a) & 900 & $\pi^{+}$ & 2 & 0.160 $\pm$ 0.002 & 0.80 $\pm$ 0.01 & 2 & 0.390 $\pm$ 0.017 & 3.924 $\pm$ 0.043 & 14.393/22 & 0.887\\
             &     & $K^{+}$   & 3 & 0.165 $\pm$ 0.004 & 0.52 $\pm$ 0.03 & 2 & 0.465 $\pm$ 0.018 & 0.480 $\pm$ 0.006 & 1.769/17 & 1\\
             &     & $p$       & 4 & 0.171 $\pm$ 0.003 & 0.51 $\pm$ 0.04 & 2 & 0.530 $\pm$ 0.020 & 0.212 $\pm$ 0.004 & 11.337/27 & 0.996\\
Figure 9 (c) &     & $\pi^{-}$ & 2 & 0.160 $\pm$ 0.002 & 0.80 $\pm$ 0.01 & 2 & 0.420 $\pm$ 0.018 & 3.884 $\pm$ 0.047 & 17.291/22 & 0.747\\
             &     & $K^{-}$   & 3 & 0.165 $\pm$ 0.004 & 0.51 $\pm$ 0.03 & 2 & 0.460 $\pm$ 0.015 & 0.472 $\pm$ 0.006 & 2.163/17 & 1\\
             &     & $\overline{p}$ & 4 & 0.168 $\pm$ 0.003 & 0.51 $\pm$ 0.02 & 2 & 0.536 $\pm$ 0.018 & 0.202 $\pm$ 0.003 & 19.958/27 & 0.833\\
\hline
Figure 9 (b) & 2760& $\pi^{+}$ & 2 & 0.156 $\pm$ 0.002 & 0.73 $\pm$ 0.01 & 2 & 0.384 $\pm$ 0.011 & 4.944 $\pm$ 0.059 & 11.582/22 & 0.965\\
             &     & $K^{+}$   & 3 & 0.190 $\pm$ 0.004 & 0.51 $\pm$ 0.04 & 2 & 0.370 $\pm$ 0.012 & 0.578 $\pm$ 0.008 & 6.135/17 & 0.992\\
             &     & $p$       & 4 & 0.179 $\pm$ 0.004 & 0.51 $\pm$ 0.04 & 2 & 0.670 $\pm$ 0.027 & 0.276 $\pm$ 0.004 & 30.949/27 & 0.273\\
Figure 9 (d) &     & $\pi^{-}$ & 2 & 0.159 $\pm$ 0.002 & 0.74 $\pm$ 0.01 & 2 & 0.387 $\pm$ 0.011 & 4.824 $\pm$ 0.058 & 16.848/22 & 0.772\\
             &     & $K^{-}$   & 3 & 0.201 $\pm$ 0.006 & 0.51 $\pm$ 0.07 & 2 & 0.330 $\pm$ 0.015 & 0.560 $\pm$ 0.008 & 9.175/17 & 0.935\\
             &     & $\overline{p}$ & 4 & 0.181 $\pm$ 0.003 & 0.51 $\pm$ 0.03 & 2 & 0.610 $\pm$ 0.024 & 0.262 $\pm$ 0.004 & 32.664/27 & 0.208\\
\hline
Figure 10 (a)&7000 & $\pi^{+}$ & 2 & 0.144 $\pm$ 0.002 & 0.60 $\pm$ 0.02 & 2 & 0.321 $\pm$ 0.007 & 6.084 $\pm$ 0.067 & 18.074/22 & 0.702\\
             &     & $K^{+}$   & 3 & 0.236 $\pm$ 0.005 & 0.62 $\pm$ 0.05 & 2 & 0.260 $\pm$ 0.012 & 0.700 $\pm$ 0.009 & 3.234/17 & 1\\
             &     & $p$       & 3 & 0.261 $\pm$ 0.006 & 0.51 $\pm$ 0.05 & 2 & 0.600 $\pm$ 0.025 & 0.346 $\pm$ 0.005 & 12.963/27 & 0.989\\
Figure 10 (c)&     & $\pi^{-}$ & 2 & 0.150 $\pm$ 0.002 & 0.63 $\pm$ 0.02 & 2 & 0.336 $\pm$ 0.007 & 5.984 $\pm$ 0.072 & 19.714/22 & 0.600\\
             &     & $K^{-}$   & 3 & 0.240 $\pm$ 0.005 & 0.63 $\pm$ 0.06 & 2 & 0.251 $\pm$ 0.014 & 0.696 $\pm$ 0.008 & 6.961/17 & 0.984\\
             &     & $\overline{p}$ & 3 & 0.260 $\pm$ 0.005 & 0.51 $\pm$ 0.05 & 2 & 0.600 $\pm$ 0.030 & 0.342 $\pm$ 0.005 & 19.058/27 & 0.868\\
\hline
Figure 10 (b)&13000& $\pi^{+}$ & 2 & 0.159 $\pm$ 0.004 & 0.70 $\pm$ 0.02 & 2 & 0.441 $\pm$ 0.027 & 5.724 $\pm$ 0.092 & 8.045/22 & 0.997\\
             &     & $K^{+}$   & 3 & 0.219 $\pm$ 0.008 & 0.71 $\pm$ 0.12 & 2 & 0.266 $\pm$ 0.031 & 0.600 $\pm$ 0.013 & 3.199/17 & 1\\
             &     & $p$       & 3 & 0.264 $\pm$ 0.009 & 0.51 $\pm$ 0.06 & 2 & 0.569 $\pm$ 0.037 & 0.314 $\pm$ 0.006 & 11.583/26 & 0.993\\
Figure 10 (d)&     & $\pi^{-}$ & 2 & 0.168 $\pm$ 0.002 & 0.73 $\pm$ 0.03 & 2 & 0.456 $\pm$ 0.039 & 5.604 $\pm$ 0.078 & 28.200/22 & 0.169\\
             &     & $K^{-}$   & 3 & 0.230 $\pm$ 0.010 & 0.69 $\pm$ 0.10 & 2 & 0.250 $\pm$ 0.026 & 0.596 $\pm$ 0.015 & 2.820/17 & 1\\
             &     & $\overline{p}$ & 3 & 0.270 $\pm$ 0.008 & 0.51 $\pm$ 0.06 & 2 & 0.569 $\pm$ 0.028 & 0.306 $\pm$ 0.006 & 10.087/26 & 0.998\\
\hline
\end{tabular}
\end{center}
}} }

The three types of yield ratios show regular trends with increase of collision energy. $k_{\pi}$ from $pp$ collisions, $k_{K}$, and $k_{p}$ increase with increase of collision energy, and $k_{\pi}$ from central Au-Au and central Pb-Pb collisions, decreases with increase of collision energy. To see more clearly the dependences of the three types of yield ratios on collision energy, we show the logarithms of the three yield ratios, $\ln(k_{\pi})$, $\ln(k_{K})$, and $\ln(k_{p})$, with $1/\sqrt{s_{NN}}$ in Figure 12. The black squares, red circles, and blue triangles denote the calculated results from inelastic $pp$, central Au-Au, and central Pb-Pb collisions at mid-rapidity, respectively. One can see that $\ln(k_{K})$ and $\ln(k_{p})$ show obviously linear dependence on $1/\sqrt{s_{NN}}$, which should be fitted by linear functions for clarity. $\ln(k_{K})$ and $\ln(k_{p})$ from all collision systems mentioned above, decrease monotonously with increase of $1/\sqrt{s_{NN}}$, and can be described by the below linear functions of
\begin{gather}
\ln(k_{K})=(-8.690\pm0.187)/\sqrt{s_{NN}}+(-0.021\pm0.030),\notag\\
\end{gather}
and
\begin{gather}
\ln(k_{p})=(-45.034\pm0.637)/\sqrt{s_{NN}}+(-0.035\pm0.030),
\end{gather}
with $\chi^2$/dof to be 7.589/26 and 3.675/17 respectively. $\ln(k_{\pi})$ displays different behavior from the above two yield ratios. With the increase of $\sqrt{s_{NN}}$, $\ln(k_{\pi})$ from inelastic $pp$ collisions increases obviously and that from nucleus-nucleus (Au-Au and Pb-Pb) collisions slightly decreases. The dependence of $\ln(k_{\pi})$ on $1/\sqrt{s_{NN}}$ can also be empirically described by the following functions of
\begin{gather}
\ln(k_{\pi_{pp}})=(-2.859\pm0.814)/\sqrt{s_{NN}}+(-0.016\pm0.045),\notag\\
\end{gather}
and
\begin{gather}
\ln(k_{\pi_{NN}})=(2.890\pm1.084)/(\sqrt{s_{NN}})^{2}+(0.205\pm0.142)/\sqrt{s_{NN}}+(-0.010\pm0.016),
\end{gather}
with $\chi^2$/dof to be 0.543/8 and 56.886/13 respectively, where $k_{\pi_{pp}}$ and $k_{\pi_{NN}}$ represent the $k_{\pi}$ from $pp$ and nucleus-nucleus (Au-Au and Pb-Pb) collisions, respectively. In fact, we also can use a linear function to fit the energy dependent $k_{\pi_{NN}}$ roughly, but in order to more accurately describe the data points in low energy region, we adopt the above polynomial function. The fitting results are represented by the solid and dotted curves, where the solid curves correspond to the data points in the energy range mentioned above, and the dotted curves show the changed trends of data points. It is noticed that the values of intercepts of the above four curves are asymptotically 0, which means the limiting values of the three yield ratios are one at very high energy.

\begin{figure}[H]
\hskip-0.0cm {\centering
\includegraphics[width=16.0cm]{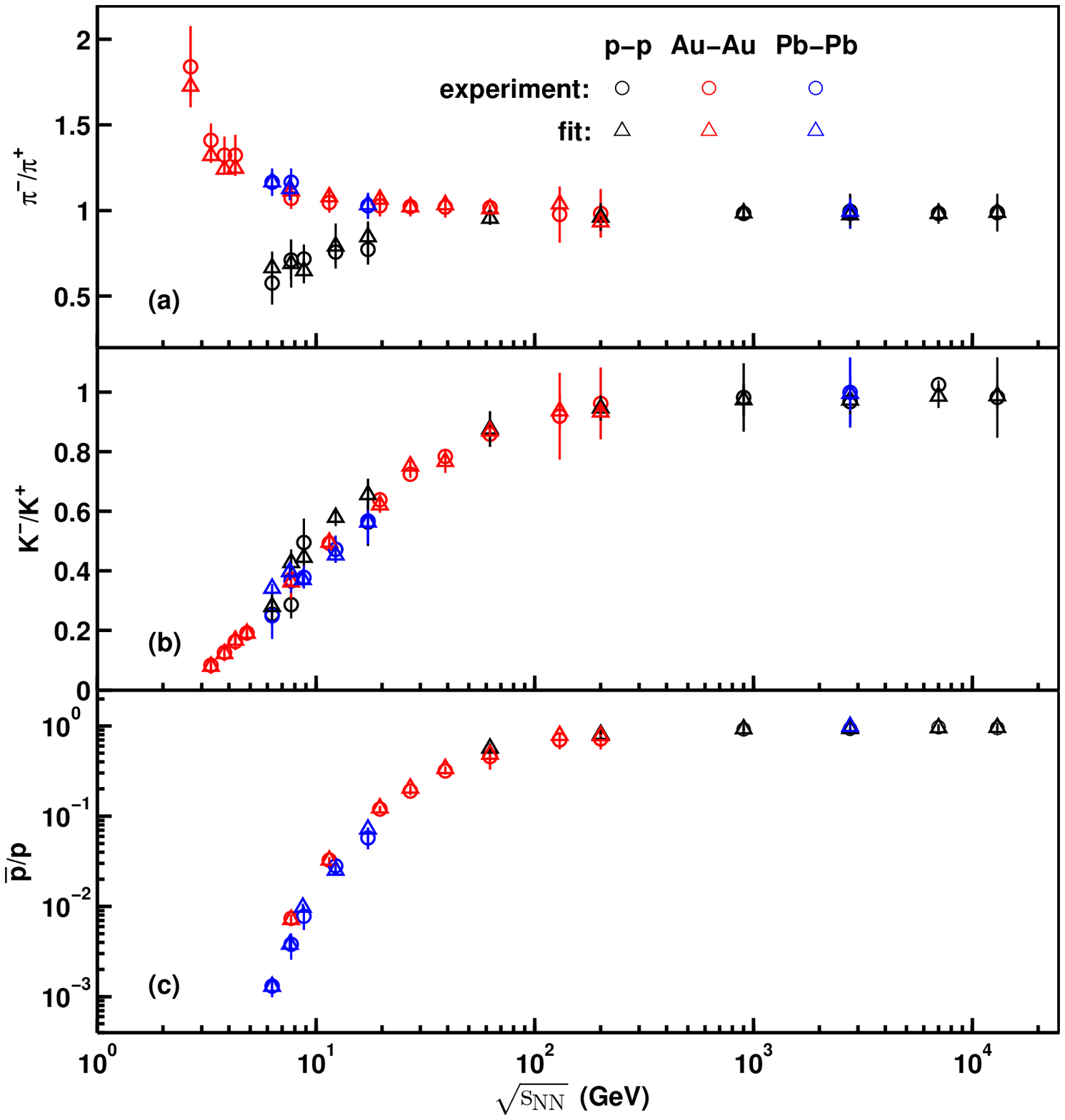}}
\vskip-0.18cm  Figure 11. $\sqrt{s_{NN}}$-dependent yield ratios, $\pi^{-}/\pi^{+}$ [Figure 11(a)], $K^{-}/K^{+}$ [Figure 11(a)], and $\overline{p}/p$ [Figure 11(c)]. The black, red, and blue circles denote respectively the calculated results from inelastic $pp$, central Au-Au, and central Pb-Pb collisions according to normalization constants, and the triangles correspondingly denote the experimental results for comparison.
\end{figure}

Based on the corrected yield ratios of negative to positive particles and Equations (8) and (9), the energy dependent chemical potentials, $\mu_{\pi}$, $\mu_{K}$, and $\mu_{p}$, of light hadrons, $\pi$, $K$, and $p$, and the chemical potentials, $\mu_{u}$, $\mu_{d}$, and $\mu_{s}$, of light quarks, $u$, $d$, and $s$, are obtained and are shown in Figure 13 with different symbols. The black squares, red circles, and blue triangles denote the calculated results from inelastic $pp$, central Au-Au, and central Pb-Pb collisions at mid-rapidity, respectively. The curves are the derivative results according to Equations (11)--(14) corresponding to the fitted curves in Figure 12. The red curves in Figures 13(a), 13(d), 13(e) and 13(f) are the derivative results related to $k_{\pi_{NN}}$ from central nucleus-nucleus collisions, and the black curves are the derivative results related to $k_{\pi_{pp}}$ from inelastic $pp$ collisions or other yield ratios. The solid and dotted curves in Figure 13 correspond to the solid and dotted curves in Figure 13 respectively. In fact, the $\mu_{p}$ in this work is close to the $\mu_{B}$ extracted from thermal fits\cite{Andronic2006,Dumitru2006,Becattini2004}, which indicates to some extent that our calculation of the chemical potential is correct. One can see that, with the increase of $\sqrt{s_{NN}}$ from AGS to LHC, $\mu_{\pi}$ from central nucleus-nucleus collisions increases obviously and that from inelastic $pp$ collisions decreases obviously, while $\mu_{K}$, $\mu_{p}$, $\mu_{u}$, $\mu_{d}$, and $\mu_{s}$ from both central nucleus-nucleus and inelastic $pp$ collisions decrease obviously. At the same energy, $\mu_{K}$ is larger than $\mu_{\pi}$ but less than $\mu_{p}$, and $\mu_{u}$ is almost as large as $\mu_{d}$ but larger than $\mu_{s}$ due to the difference in mass between different particles. The limiting values of the six types of chemical potentials from central nucleus-nucleus and inelastic $pp$ collisions are zero at very high energy.

\begin{figure}[H]
\hskip-0.0cm {\centering
\includegraphics[width=16.0cm]{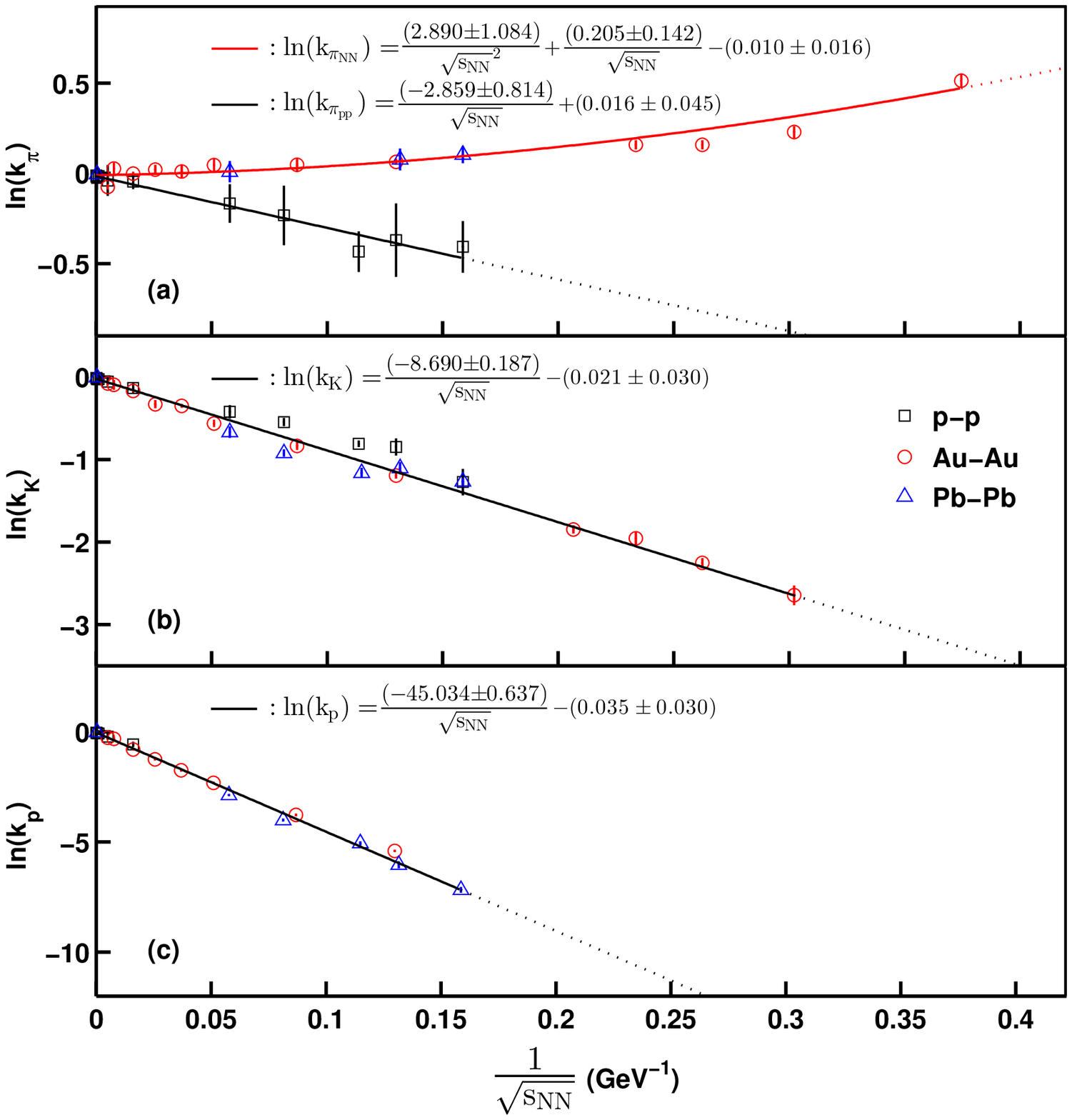}}
\vskip-0.18cm  Figure 12. Energy dependent and modified yield ratios. The symbols denote the yield ratios modified by removing strong decay from high-mass resonance and weak decay from heavy flavor hadrons. The curves are the fits according to Equations (11)--(14), respectively.
\end{figure}

From Figure 12 one can see that, in central Au-Au and Pb-Pb collisions, $\ln(k_{\pi_{NN}})$ $(>0)$ increases and $\ln(k_{K})$ $(<0)$ and $\ln(k_{p})$ $(<0)$ decrease obviously with the increase of energy. The difference between $\ln(k_{\pi_{NN}})$ and $\ln(k_{K})$ ($\ln(k_{p})$) is caused by different production mechanisms. In the processes of producing pions, kaons, and protons, the difference of cross-section of absorbtion, content of primary proton in nuclei and so on can result in the difference in the yield of these particles. $\ln(k_{\pi_{NN}}) > 0$ and $\ln(k_{\pi_{pp}}) < 0$ result in $\mu_{\pi_{NN}} > 0$ and $\mu_{\pi_{pp}} < 0$. As the energy increases to the LHC, in both central Au-Au (Pb-Pb) collisions and $pp$ collisions, $k_{\pi}$, $k_{K}$, and $k_{p}$ approach to one, and $\mu_{\pi}$, $\mu_{K}$, and $\mu_{p}$ approach to zero. These same limiting values indicate that hard scattering process possibly plays an important role, mean-free-path of particles becomes largely, and the collision system possibly changes completely from the hadron-dominant state to the quark-dominant state.

In Figure 13, it should be noted that the derived curves of hadron and quark chemical potentials from the fits of the energy dependent yield ratios in Figure 12 simultaneously show a maximum at around 4 GeV, which is not observed from the fits of yield ratios. In order to figure out the accurate energies at these maximums, we make the following calculation according to Equations (6), (8), (9), (11)--(14). For these black curves, whose derivation does not involve Equation (14), we figure out the analytical solutions for the energies at these maximums. The derivation is as follows. From the fits of the logarithms of the three types of yield ratios with $1/\sqrt{s_{NN}}$ in Equations (11)--(13), one can see that all the intercepts on the vertical axes approximate to zero. For simplicity of calculation, we assume that all the intercepts are zero, then Equations (11)--(13) have the following form of
\begin{equation}
\ln(k_{i})=\frac{A_i}{\sqrt{s_{NN}}} \quad   (i=\pi, K, and \,\, p),\\
\end{equation}
where $A_i$ is the slope. Then, according to Equations (8) and (9), the chemical potential $\mu_j$ can be given by
\begin{equation}
\mu_j=T_{ch}\frac{B_j}{\sqrt{s_{NN}}}  \quad   (j=\pi, K, p, u, d, and \,\, s),\\
\end{equation}
where $B_j$ is a constant. In consideration of $T_{ch}$ given by Equation (6), $\mu_j$ can be written as
\begin{equation}
\mu_j=T_{\lim}B_j\{[1+\exp(2.60-\ln{\sqrt{s_{NN}}}/0.45)]\sqrt{S_{NN}}\}^{-1}.  \\
\end{equation}
Then,
\begin{equation}
\frac{\mathrm{d}\mu_j}{\mathrm{d}\sqrt{S_{NN}}}=T_{\lim}B_j\{[1+\exp(2.60-\ln\sqrt{s_{NN}}/0.45)]\sqrt{S_{NN}}\}^{-2}[1-({11}/{9})\exp(2.60-\ln\sqrt{s_{NN}}/0.45)].\\
\end{equation}
Let $\frac{\mathrm{d}\mu_j}{\mathrm{d}\sqrt{s_{NN}}}=0$, then,
\begin{equation}
1-(11/9)\exp(2.60-\ln\sqrt{s_{NN}}/0.45)=0.
\end{equation}
Finally, we obtain the energy value at the maximum, $\sqrt{s_{NN}}=3.526$ GeV. One can see that all the energy values at these maximums are the same, which means that in the case of linear fitting (Equations (11)--(13)), the energy is independent of the slope parameter of linear equation. It needs to be emphasized that, in Figure 12, due to the lack of data in low energy region, we can only make prediction about them by the linear fits that can well describe the data points in relatively high energy region. That means the obtained energy value at the maximum is based on the above linear fits. For these red curves, whose derivation involves Equation (14), due to the calculation being complicated, we only give the numerical solutions for the energies at the maximums of these curves. From the data of these derived curves obtained by Equations (6), (8), (9), (11)--(13) in Figure 13, we give the numerical solutions for the energies at the maximums of all curves. From Figure 13(a) to Figure 13(f), the values of the energies at these maximums one by one are 3.584 GeV (for black curve), 3.398 GeV (for red curve), 3.310 GeV, 3.534 GeV, 3.537 GeV (for black curve), 3.590 GeV (for red curve), 3.527 GeV (for black curve), 3.428 GeV (for red curve), 3.521 GeV (for black curve), 3.666 GeV (for red curve). One can see that the energies at these maximum range from 3.310 Gev to 3.666 Gev, and the average value of these energies is 3.510 GeV.

\begin{figure}[H]
\hskip-0.0cm {\centering
\includegraphics[width=16.0cm]{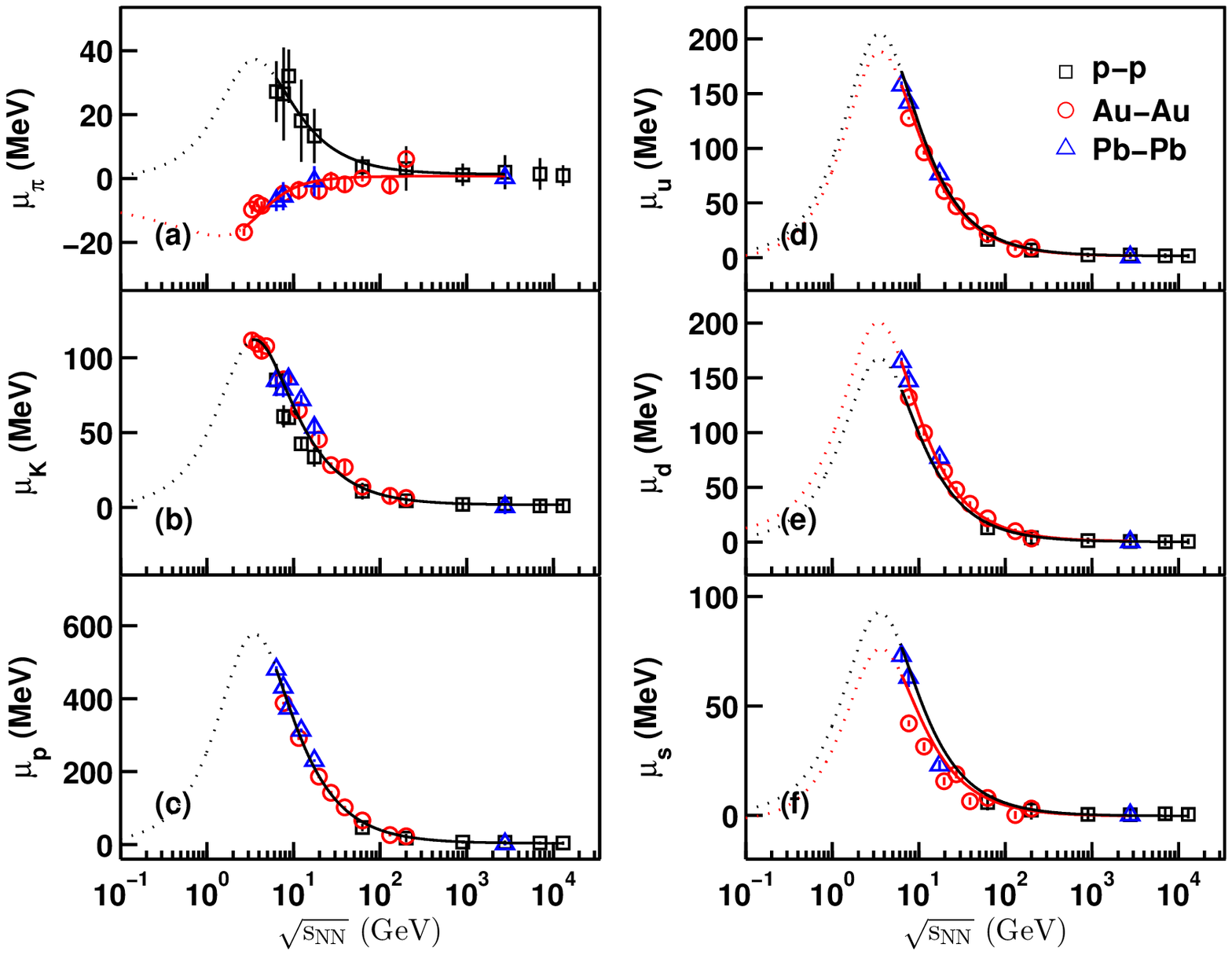}}
\vskip-0.18cm  Figure 13. Energy dependent chemical potentials, $\mu_{\pi}$ [Figure 13(a)], $\mu_{K}$ [Figure 13(b)], and $\mu_{p}$ [Figure 13(c)], of light hadrons, $\pi$, $K$, and $p$, and $\mu_{u}$ [Figure 13(d)], $\mu_{d}$ [Figure 13(e)], and $\mu_{s}$ [Figure 13(f)], of light quark, $u$, $d$, and $s$. The symbols denote the calculated results according to the modified yield ratios and Equations (8) and (9), and the curves are the derivative results based on the fits of Equations (11)--(14) in Figure 12.
\end{figure}

The special energy (around 3.510 GeV) at the maximum possibly is the critical energy of phase transition from a liquid-like hadron state to a gas-like quark state in the collision system, where the liquid-like state and the gas-like state are the states in which the mean-free-path of interacting particles are relatively short and relatively long, respectively. In other words, at this special energy, the collision system starts to change initially its state from the liquid-like nucleons and hadrons to the gas-like quarks, and many properties of the system also change. The curve of proton chemical potential having the maximum at this special energy, indicates that the density of baryon number in nucleus-nucleus collisions has the largest value and the mean-free-path of particles has the smallest value at this special energy, which means that the hadronic interactions play an important role at this stage \cite{Braun-Munzinger2007}. When $\sqrt{s_{NN}}$ > 3.510 GeV, the chemical potential gradually decrease with increasing $\sqrt{s_{NN}}$, implying that the density of baryon number gradually decreases \cite{Braun-Munzinger2007},  shear viscosity over entropy density gradually weakens \cite{Gyulassy2005}, and mean-free-path gradually increases. Meanwhile, the hadronic interactions gradually fade and the partonic interactions gradually become greater. When $\sqrt{s_{NN}}$ increases to the top RHIC, especially the LHC, all types of chemical potentials approach to zero, when the high energy collision system possibly changes completely from the hadron-dominant state to the quark-dominant state and signifies that the partonic interactions possibly play a dominant role at the top RHIC and LHC \cite{Xu2014,Abelev(STARCollaboration)2009paper1}, and the strongly coupled QGP (sQGP) has been observed  \cite{Bazavov2012,Adams(STARCollaboration)2005,Grosse-Oetringhaus(theALICECollaboration)2014}. It should be pointed out that, due to the lack of experimental data in low energy range, the existence of this maximum is not actually certain. The maximum is only a calculated result according to some empirical formulas, so the energy of critical point is large fluctuations. Although the trend of the chemical potential in low energy region ($\sqrt{s_{NN}}$ < 3.510 GeV) is unlikely, the maximum point is at least a turning point, which implies the possibility of phase transition. These results are consistent with our previous work \cite{He2019}. Our result (around 3.510 GeV) of the critical energy of phase transition is consistent with that (below 19.6 GeV) by the STAR Collaboration \cite{Xu2014}, and less than the result (between 11.5 GeV and 19.6 GeV) of a study based on the correlation between collision energy and transverse momentum \cite{Abelev(ALICECollaboration)2015,Md.Nasim2015,Abelev(STARCollaboration)2009paper1} and the result (around 62.4 GeV) of the study based on a striking pattern of viscous damping and an excitation function \cite{Lacey2015}. One can see that although there are many study results, none of them have been confirmed to be reliable. Therefore, we need to continue to study and confirm the exact critical energy of the phase transition.

{\section{Summary and Conclusion}}

The transverse momentum (or mass) spectra of final-state light flavour hadrons, $\pi^{\pm}$, $K^{\pm}$, $p$, and $\bar p$, produced in central Au-Au, central Pb-Pb and inelastic $pp$ collisions at mid-rapidity over an energy range from AGS to LHC, are described by a two- or one-component Erlang distribution in the frame of multi-source thermal model. The fitting results are in agreement with the experimental data recorded by the E866, E917, E895, NA49, NA44, NA61/SHINE, PHENIX, STAR, and ALICE Collaborations.

From the fitting parameters, in most cases, the experimental data of $p_{T}$ (or $m_{T}$) spectra are suitable for the two-component Erlang distribution, where the first component corresponding to a narrow low-$p_{T}$ (or $m_{T}$) region is contributed by the soft excitation process in which a few sea quarks and gluons take part in, and the second component corresponding to a wide high-$p_{T}$ (or $m_{T}$) region is contributed by the hard scattering process which is a more violent collision between two valent quarks in incident nucleons. The study shows that the values of the contribution ratio of soft excitation process are more than 50\%, which means that the excitation degrees of these collision systems are mainly contributed by the soft excitation processes.

Based on the normalization constants in fitting the transverse momentum or mass spectra of final-state light flavour particles, the final-state yield ratios of negative to positive particles are obtained. The energy dependent chemical potentials of light hadrons, $\mu_{\pi}$, $\mu_{K}$, and $\mu_{p}$, and quarks, $\mu_{u}$, $\mu_{d}$, and $\mu_{s}$, are extracted from the modified yield ratios in which the contributions of strong decay from high-mass resonance and weak decay from heavy flavor hadrons are removed. With the increase of $\sqrt{s_{NN}}$ over a range from a few GeV to more than 10 TeV, the $\mu_{K}$, $\mu_{p}$, $\mu_{u}$, $\mu_{d}$, $\mu_{s}$ decrease obviously in central Au-Au, central Pb-Pb, and inelastic $pp$ collisions, while $\mu_{\pi}$ increases in central Au-Au and Pb-Pb collisions and it decreases in inelastic $pp$ collisions. When collision energy increases to the top RHIC and LHC, all types of chemical potentials are small and the limiting values of them are zero in central Au-Au, central Pb-Pb, and inelastic $pp$ collisions at very high energy.

The logarithms of the yield ratios, $\ln(k_{\pi_{pp}})$, $\ln(k_{K})$, and $\ln(k_{p})$, show obviously linear dependences on $1/\sqrt{s_{NN}}$ in mentioned energy range, and $\ln{k_{\pi_{NN}}}$ increases as a polynomial function of $1/\sqrt{s_{NN}}$. Base on the above relationships, we find that at about 3.510 GeV, the derived curves of hadron and quark chemical potentials simultaneously show the maximum. This special energy possibly is the critical energy of phase transition from a liquid-like hadron state to a gas-like quark state in high energy collision system, where the density of baryon number in nucleus-nucleus collisions has a large value and the hadronic interactions play an important role at this stage. When collision energy increases to the top RHIC, especially the LHC, all types of chemical potentials approach to zero, which indicates that hight energy collision system possibly changes completely from the hadron-dominant liquid-like state to the quark-dominant gas-like state and the partonic interactions possibly play a dominant role at the LHC. \\

{\bf Acknowledgments}

This work was supported by the National Natural Science Foundation of China under Grant No. 11847114.\\

{\bf Data availability}

All data are quoted from the mentioned references. As a phenomenological work, this paper does not report new data.\\

{\bf Conflicts of Interest}

The authors declare that there are no conflicts of interest regarding the publication of this paper.\\

\end{document}